# Evaluating the Performance of Low-Cost PM$_{2.5}$ Sensors in Mobile Settings

Priyanka deSouza[1,2*], An Wang[3], Yuki Machida[3], Tiffany Duhl[4], Simone Mora[3], Prashant Kumar[5,6], Ralph Kahn[7], Carlo Ratti[3], John L Durant[4], Neelakshi Hudda[4]

1: Department of Urban and Regional Planning, University of Colorado Denver, CO 80217, USA
2: CU Population Center, University of Colorado Boulder, Boulder CO 80309, USA
3: MIT Senseable City Lab, Cambridge MA 02139, USA
4: Department of Civil and Environmental Engineering, Tufts University, Medford, MA 02155, USA
5: Global Centre for Clean Air Research (GCARE), School of Sustainability, Civil and Environmental Engineering, Faculty of Engineering and Physical Sciences, University of Surrey, Guildford GU2 7XH, Surrey, UK
6: Institute for Sustainability, University of Surrey, Guildford GU2 7XH, Surrey, UK
7: NASA Goddard Space Flight Center, Greenbelt MD 20771, USA

*Corresponding author: priyanka.desouza@ucdenver.edu

## Abstract

Low-cost sensors (LCS) for measuring air pollution are increasingly being deployed in mobile applications but questions concerning the quality of the measurements remain unanswered. For example, what is the best way to correct LCS data in a mobile setting? Which factors most significantly contribute to differences between mobile LCS data and higher-quality instruments? Can data from LCS be used to identify hotspots and generate generalizable pollutant concentration maps? To help address these questions we deployed low-cost PM$_{2.5}$ sensors (Alphasense OPC-N3) and a research-grade instrument (TSI DustTrak) in a mobile laboratory in Boston, MA, USA. We first collocated these instruments with stationary PM$_{2.5}$ reference monitors (Teledyne T640) at nearby regulatory sites. Next, using the reference measurements, we developed different models to correct the OPC-N3 and DustTrak measurements, and then transferred the corrections to the mobile setting. We observed that more complex correction models appeared to perform better than simpler models in the stationary setting; however, when transferred to the mobile setting, corrected OPC-N3 measurements agreed less well with corrected DustTrak data. In general, corrections developed using minute-level collocation measurements transferred better to the mobile setting than corrections developed using hourly-averaged data. Mobile laboratory speed, OPC-N3 orientation relative to the direction of travel, date, hour-of-the-day, and road class together explain a small but significant amount of variation between corrected OPC-N3 and DustTrak measurements during the mobile deployment. Persistent hotspots identified by the OPC-N3s agreed with those identified by the DustTrak. Similarly, maps of PM$_{2.5}$ distribution produced from the mobile corrected OPC-N3 and DustTrak measurements agreed well. These results suggest that identifying hotspots and developing generalizable maps of $PM_{2.5}$ are appropriate use-cases for mobile LCS data.

# 1 Introduction

Exposure to fine particulate matter ($PM_{2.5}$; mass concentration of particles having an aerodynamic diameter of ≤2.5 µm) is a major environmental health risk in cities around the world[1–6]. Air quality monitoring is critical for developing effective pollution management plans; however, reference air quality monitors have high capital and operating costs. Even in the United States, which has one of the densest monitoring networks in the world, less than a third of US counties have a regulatory, also termed reference, monitor[7]. Such a sparse network cannot adequately characterize the spatial variation of air pollution in complex urban environments where pollutant concentrations can vary significantly at distances as small as a few 10s of meters[8,9].

Dense networks of low-cost sensors (LCS) (<USD $2500 as defined by the US EPA Air Sensor Toolbox[10]) are increasingly being used to fill $PM_{2.5}$ concentration data gaps in cities around the world[11–16]. In particular, there is growing interest in using LCS in mobile monitoring applications to capture the granular spatial variations in pollution in cities at a relatively low-cost. However, the measurements from LCS are error-prone and need to be corrected prior to data interpretation[17–20].

Due to the growing use of LCS, regulatory agencies in the United States and Europe have developed protocols to evaluate LCS in the lab and in the field. Although these protocols offer an important point of reference to evaluate LCS, considerable challenges remain in using them. For example, the agreement of LCS measurements with those from reference monitors varies with relative humidity (RH), temperature (T), as well as aerosol concentration, size distribution, and composition. Mobile deployments of LCS may introduce additional uncertainties due to vibrations, wind speed and vehicular disturbances of $PM_{2.5}$ mixing patterns. Therefore, LCS need to be systematically evaluated against reference monitors to appropriately correct high-frequency mobile measurements as well as to assess appropriate use-cases for mobile LCS data.

To date, little research has evaluated LCS in mobile settings. We found only one study that systematically evaluated the performance of LCS for $PM_{2.5}$ (PurpleAir PA-II monitors) against measurements from collocated reference instruments in a mobile platform[21]. This study found that the fidelity of the PurpleAir measurements decreased with increasing air velocity around the unit.

We were interested in evaluating the transferability of correction models developed based on a stationary collocation experiment to a mobile deployment of a popular LCS, the Alphasense Optical Particle Counter (OPC-N3). The impact of correction model choice on the performance of an LCS in a mobile setting has not previously been reported in the literature. In addition, we evaluate the performance of corrected Alphasense OPC-N3s under different driving conditions and in different urban microenvironments. We report on which factors (i.e., sensor orientation, vehicle speed, road class, and meteorology) contribute to differences between the corrected OPC-N3 and $PM_{2.5}$ measurements made using a research grade monitor (TSI DustTrak). Finally, we evaluate the feasibility of using mobile LCS for two common use-cases: detecting $PM_{2.5}$ hotspots and producing generalizable maps of $PM_{2.5}$ concentrations over a sampling area.

# 2 Materials and methods

## 2.1 City Scanner: A low-cost sensor suite

As part of the City Scanner project at the Senseable City Lab, Massachusetts Institute of Technology, we developed a customizable low-cost sensor suite for mobile applications[22,23]. For this study, we used two City Scanner sensor suites each equipped with an Alphasense OPC-N3[24] optical particle counter to measure particulate matter, a SHT20[25] to measure temperature and humidity, a GPS sensor, and flash memory. Data was transmitted to a server via the Global System for Mobile Communication (GSM) and accessed remotely.

### 2.1.1 Low-cost $PM_{2.5}$ sensors: Alphasense OPC-N3

The OPC measures the intensity of light scattered by each particle as sampled air is illuminated by a laser beam. The amount of light scattered is a function of particle size. The manufacturer uses monodisperse particles of known size to calibrate the OPC, deriving counts in 24 particle-size bins ranging from 0.35 to 40 μm. An Alphasense data reduction algorithm converts the particle counts to $PM_{2.5}$ mass concentration. A brief overview of past evaluations of the OPC-N3 is provided in **section S1** of the *Supplementary Information (SI)*. We used a several-years-old OPC-N3 (referred to as OPC 1) and a new OPC-N3 (referred to as OPC 2) in this experiment.

## 2.2 High-cost portable $PM_{2.5}$ monitor: TSI DustTrak™

We used a TSI DustTrak II 8533 as a reference instrument for the mobile monitoring experiments. The DustTrak uses a laser photometer to determine real-time PM mass concentration based on $90^0$ light scattering. The DustTrak II 8533[26] detects particles in the size range from 0.1 to 15 μm in diameter, and has a mass concentration resolution of ±0.1% or 1 μg/m$^3$ (whichever is greater). DustTraks report $PM_1$, $PM_{2.5}$, $PM_4$, and $PM_{10}$. We selected the DustTrak because it has a high reported precision[27].

## 2.3 Experimental Design

We collocated the two CityScanners at a stationary monitoring site containing a reference monitor and we developed correction models (also referred to as calibration models) for the OPC-N3s using the reference measurements as the gold standard. We separately collocated a DustTrak monitor with a reference monitor in another stationary reference monitoring site and developed a correction model for the DustTrak. We then mounted the two CityScanners and the DustTrak in our mobile laboratory and collected measurements while driving prescribed routes in Boston. We evaluated the transferability of different OPC-N3 correction models to the mobile setting by comparing corrected OPC-N3 measurements with corrected DustTrak data. We also evaluated how much of the variation between the corrected mobile OPC-N3 and DustTrak measurements were due to factors such as orientation of the OPC-N3, vehicle speed, sky view factor (fraction of visible sky that provides an indication of street geometry and building density[28]), meteorology, time-of-day and day. We then evaluated the performance of the corrected mobile OPC-N3 measurements vis a vis the DustTrak for two use-cases.

## 2.3.1 Performance evaluation in a stationary setting under well-characterized ambient conditions

We collocated two Alphasense OPC-N3s and a reference monitor between April 11, 2022 11:40 AM and April 27, 2022 4:36 PM at the EPA Speciation Trends Network reference site (Site ID:25-025-0042; maintained by the Massachusetts Department of Environmental Protection (MADEP)) in Nubian Square, Boston, which is representative of urban background. This site is within the study area covered by our mobile monitoring campaign.

We collocated the DustTrak at a MADEP reference site located adjacent to Interstate-93 in Boston (Site ID: 25-025-0044; also within the study area) between June 29, 2022, 8:50 AM and July 5, 2022, 8:30 AM. The reference monitor at both MADEP sites was a Teledyne T640 monitor that had a sampling frequency of 1 minute.

The sampling frequencies of the OPC-N3s and DustTrak during stationary collocation were 5 seconds and 1 minute, respectively. We averaged the OPC-N3 data every 1 minute and merged the minute-averaged data with coincident 1-minute measurements from the reference monitor. Similarly, we merged the DustTrak measurements with coincident reference measurements. We also averaged the OPC-N3 and DustTrak measurements every hour and merged the hourly-averaged data with coincident hourly-averaged measurements from the reference monitors. We had a total of 22,882 minute-averaged, and 386 hourly-averaged coincident OPC and reference measurements. We had a total of 8,618 minute-averaged, and 143 hourly-averaged coincident DustTrak and reference measurements.

We evaluated the two OPC-N3s performance vis-a-vis each other using two metrics: (1) Pearson Correlation Coefficient (R), and (2) the root-mean-square-error (RMSE). We chose to also evaluate the RMSE because R might be a misleading indicator of sensor performance when $PM_{2.5}$ concentrations are close to the limit of detection of the instruments. In addition, we compared the sensor measurements in the stationary setting after using empirical corrections (described in the next section) to correct the OPC-N3 and DustTrak $PM_{2.5}$ measurements using the reference measurements.

### 2.3.1.1 Defining the Models Used to Correct the OPCs

The correction models were developed to predict the true $PM_{2.5}$ concentrations as accurately as possible. The reference measurements, considered as *true* $PM_{2.5}$ concentrations, were the dependent variable in the models.

We evaluated 17 correction models that included different combinations of temperature (T), relative humidity (RH) and dewpoint (D) (**Table S1**). We used T and RH reported by each CityScanner mobile sensor suite as it most closely represents the conditions experienced by the OPCs. R obtained when comparing minute-averaged T and RH experienced by the collocated OPCs were both ~0.95. The R observed when comparing minute-averaged T and RH from the OPCs with that reference site were 0.92 and 0.90, respectively (**Figure S1**). We derived (D) from T and RH reported by the CityScanner sensor suite using the *weathermetrics* package in the programming language R[29], as D has been shown to be a good proxy of particle hygroscopic growth[30,31].

Sixteen of the 17 correction models were multivariate regression models that were also applied in recent papers[18,30] to correct networks of low-cost sensors. As T, RH, and D are not independent (**Figure S2**), the 16 linear regression models include adding the meteorological conditions as interaction terms, instead of additive terms. The seventeenth correction model relies on the machine learning technique: Random Forest (RF). Machine learning models can capture more complex nonlinear effects, such as unknown relationships between additional spatial and temporal variables. We opted to use the RF as it has been widely used in correcting LCS[32,33]. A description of the RF technique is described in detailed in *SI* **Section S2**. We used a 10-fold cross-validation (CV) method to avoid overfitting in the RF model. For more details on the cross-validation methods refer to *SI* **Section S3**.

We derived separate correction models for each OPC for minute-averaged and hourly-averaged stationary-site and used the models to correct the mobile OPC data. We report the R and RMSE when comparing predicted measurements derived using each correction model with the corresponding reference measurements.

### 2.3.1.2 Correcting the DustTrak

We applied three widely-used correction methods[34] to correct the DustTrak data, including a simple bias correction, a RH-adjusted empirical equation, and an RH-adjusted linear regression model described in *SI* **Section S4**. We evaluated the R and RMSE to compare corrected DustTrak measurements using each of these three methods with corresponding reference measurements.

We applied the correction developed using minute-averaged data to the DustTrak data in the mobile setting, as this frequency best matches the frequency of measurements of the mobile (5 s). In this experiment, we use corrected DustTrak data as our gold standard for the mobile deployment. In supplementary analyses, we used the correction developed using hourly-averaged collocation data to correct DustTrak measurements in the mobile setting.

## 2.3.2 Performance evaluation of the OPC measurements in a mobile setting

We evaluated the transferability of each correction technique for the OPCs to a mobile setting by examining if there was an improvement in R and RMSE in comparing the corrected versus uncorrected OPC measurements with the coincident corrected DustTrak data. We also evaluated if the distribution of T and RH during the collocation period were representative of the mobile deployment (*SI* **Figures S3, S4**).

We collocated the two CityScanners alongside the DustTrak in a mobile laboratory[35]. The mobile laboratory was an electric vehicle with a battery-powered electricity supply for the DustTrak; thus obviating the need for measurement correction due to self-sampling of emissions from the vehicle; it is described in detail elsewhere[35]. The DustTrak was placed inside the mobile laboratory and plumbed into the main plenum via a 45-cm-long conductive tube. The two City Scanners were placed atop the mobile lab.

City Scanner 1 containing OPC 1 was deployed for ~ 58 hours over 12 unique dates between Feb 11, 2022 and April 5, 2022. City Scanner 2 containing OPC 2 was deployed for ~ 14 hours over 3 unique dates between March 23, 2022 and April 5, 2022. The orientation of the inlet of OPC 1 relative to the direction of the vehicle (parallel or perpendicular) was varied during the course of the experiment (**Table S2**). The sampling routes were selected to encounter a wide range of microenvironments such as highways, tunnels, and commercial and residential streets.

The sampling frequencies of the OPC-N3s and the DustTrak were 5 s. We had a total of 34,127 coincident OPC 1 and DustTrak mobile measurements over 12 separate mobile monitoring runs, and 7,972 coincident OPC 1, OPC 2 and DustTrak measurements over 3 separate mobile monitoring runs. We present descriptive statistics of each run (**Table S2**). Maps of the sampling route for each run are displayed in *SI* **Figure S5**.

We compared corrected mobile OPC measurements using the different techniques proposed in **Table S1**, and corrected DustTrak measurements (**Table 1**). We report which correction model transferred best to the mobile setting. In *SI* **Table S3** in *Supplementary Information*, we compare corrected OPC and DustTrak measurements using mobile DustTrak data corrected using hourly-averaged collocated measurements in the stationary setting.

We carried out an Analysis of Variance (ANOVA) test to determine which factors most explained the absolute difference, and difference between the corrected and uncorrected OPC measurements, and the corrected OPC and DustTrak measurements. We considered RH, T, time-of-day, day-of-run, vehicle speed, inlet orientation, spatial covariates such as class of road the measurement was made on, and skyview factor (SVF) due to buildings in the 100-m radius around each mobile measurement location. For 1,529 (4.5%) and 3875 (11.4%) mobile OPC 1 measurements, we did not have SVF and road class data, respectively. For 308 (3.9%) and 1,533 (19.2%) mobile OPC 2 measurements, we did not have SVF and road class data, respectively. Given the large number of measurements missing road class information, we did not include road class in our main analysis, but instead included it in supplementary analyses (*SI* **Table S4**).

## 2.4 Evaluating use-cases of mobile LCS

We evaluated two use-cases. First, we studied the identification of hotspots and overlap in persistent hotspots as measured by low-cost versus research-grade instruments. Hotspots were defined as the top one percentile of raw $PM_{2.5}$ concentrations measured by each instrument. Persistent hotspots were defined as $PM_{2.5}$ measurements in the top one percentile of all measurements in the same location over multiple days. Hierarchical clustering was used to identify hotspots from the highest $PM_{2.5}$ concentrations reported by each instrument in the same spatial area using a distance cut-off of 100 m. Thereupon we determined the number of measurements (n) in each cluster, as well as the number of unique runs (nr) over which measurements in each cluster were made, and lastly defined a persistent hotspot as a cluster for which n > 5 and nr > 1. Uncorrected $PM_{2.5}$ measurement from OPC 1 were compared to DustTrak measurements to evaluate this use-case, as previous work has indicated that raw $PM_{2.5}$ measurements from low-cost sensors can be used to detect hotspots[8]. In this analysis, we used only measurements for OPC 1 because it was in operation for 12 days as opposed to 3 days for OPC 2.

The second use-case was the development of 'generalizable' maps of $PM_{2.5}$ concentrations for the study domain. The maps are produced according to the following steps:

(1) Calculating a background correction to compare $PM_{2.5}$ measurements made at the same location but on different days and at different times.
(2) Applying the background correction to the $PM_{2.5}$ concentrations from each device.
(3) Dividing the sampling area into 50 m x 50 m grid cells and then calculating the median background-corrected $PM_{2.5}$ concentration for each grid cell in the study-area. We opted to use a 50 m resolution as it is an order of magnitude greater than the GPS error (6 m), but small enough to derive granular maps of $PM_{2.5}$ concentrations for the study area.

We evaluated the median generalizable $PM_{2.5}$ concentrations, and sampling error, for each grid cell in the sampling area (3,824 grid cells had OPC 1 measurements, 1,763 grid cells had OPC 2 measurements) for the corrected and uncorrected data from the OPCs and the corrected measurements from the DustTrak. Section **S5** in the *SI* contains more information about how such 'generalizable' $PM_{2.5}$ concentrations and sampling errors were calculated. We examined correlations between the maps of generalizable $PM_{2.5}$ concentrations produced across (1) all 30 m road segments, (2) only segments where measurements were made over more than a single run and the normalized standard error in the generalizable $PM_{2.5}$ concentrations calculated from the sampling variability of each instrument was < 20%.

All analyses were conducted using the software R. In all analyses, p-values < 0.05 were taken to represent statistical significance.

# 3 Results and Discussion

During the collocation, RH spanned the range of conditions observed in the mobile deployment, although some temperatures during the mobile deployment were higher (> $25^0$C) than during the stationary collocation (≤ $25^0$C) (**Figures S3-4**).

## 3.1 Evaluating Corrections Developed During the Collocation

The R and RMSE obtained when comparing the uncorrected, minute-averaged DustTrak and reference monitoring data were 0.81 and 7.8 µg/m³, respectively. When repeating this calculation using hourly-averaged uncorrected data we obtained an R and RMSE of 0.77 and 8.8 µg/m³, respectively. Overall, we found that a simple bias correction of data from the DustTrak (Equation 1 in *SI* **Section S4**) resulted in the best performance, yielding an R and RMSE of 0.77 and 3.3 µg/m³, and 0.81 and 2.0 µg/m³ using hourly- and minute-averaged data, respectively.

As expected, the uncorrected measurements from the newer OPC, OPC 2, were more strongly correlated with reference measurements than that of the older OPC, OPC 1. When comparing OPC measurements corrected using the different correction functions outlined in **Table S1** (and visualized in **Figure S6**) with that of the corresponding reference PM$_{2.5}$ data in the stationary setting, we observed that the RF model yielded the best performance (R > 0.9, RMSE < 1 µg/m$^3$) using both hourly- and minute-averaged data for both OPCs. More complex linear regression models (Models 13, 14, 15 and 16) outperformed simpler models (Models 1 - 12) (**Table S1**).

## 3.1 Evaluating transferability of OPC corrections developed in a stationary setting to the mobile deployment

For corrections developed using minute-averaged collocation data, although the RF model yielded the best performance in the stationary setting, it yielded a worse performance (R= 0.17 for OPC 1 and R = 0.16 for OPC 2) than even the uncorrected OPC measurements (R = 0.26 for OPC 1 and R =0.20 for OPC 2) when comparing OPC and DustTrak measurements in the mobile setting (**Table 1**; visualized in **Figure S7**). For OPC1, more complex linear regression models (Models 10 -16) also yielded worse R and RMSE when applied to the mobile setting than uncorrected OPC data. For OPC 2, Models 10 to 15 yielded a small improvement in performance relative to the raw data.

For both OPCs, **Model 6** (which corrected for RH, R and the interaction term between RH and D) yielded the largest improvement in R and RMSE (R and RMSE comparing OPC 1 and the DustTrak were 0.29 and 5.6 µg/m$^3$, R and RMSE comparing OPC 2 with the DustTrak were 0.59 and 3.9 µg/m$^3$), and also led to the best fit between the corrected measurements of the two OPCs themselves (R and RMSE were 0.65 and 1.4 µg/m$^3$). The higher correlation observed between OPC 2 and the DustTrak is likely since OPC 2 was a newer OPC and was not degraded.

We obtained similar findings when evaluating the transferability of hourly-averaged correction models (**Table 1**; visualized in **Figures S8 and S9**). **Model 6** resulted in the biggest improvement of R for both OPCs in the mobile setting. However, for correction models developed using hourly-averaged stationary OPC data, it appears that most corrections resulted in a dramatic increase in RMSE for OPC 1, compared to using uncorrected measurements (**Table 1**). This is likely because the hourly-averaged data used to develop the bias correction of OPC 1 did not include the sharp increases in PM$_{2.5}$ we observed in minute-averaged measurements during the stationary collocation, which were also observed during the mobile experiment. (**Figures S10** and **S11** display collocated OPC 1 and reference data at the minute- and hourly-averaged resolutions, respectively; **Figures S12** and **S13** displays collocated OPC 2 and reference data at the minute- and hourly-averaged levels, respectively). Thus, the corrections developed using hourly-averaged measurements under-correct OPC 1. The spikes observed in the OPC 2 PM$_{2.5}$ concentrations in the stationary setting were smaller, likely because it is a new OPC and is less error (spike)-prone. However, for both OPCs, using minute-averaged data at the reference site to correct the raw OPC data across all models yielded the best comparison between the corrected OPC and DustTrak measurements.

We repeated the exercise of comparing corrected-OPC and DustTrak measurements with the DustTrak data corrected using hourly-averaged $PM_{2.5}$ concentrations instead of minute-averaged measurements from the collocation in supplementary analyses (**Table S3**). **Model 6** developed using minute-averaged OPC collocation data resulted in the most striking improvement in R and RMSE when comparing corrected OPC 2 and DustTrak data. **Model 6** also resulted in a smaller improved performance for OPC 1 (**Model 4** yielded a marginally smaller RMSE). Overall, this supplementary analysis confirmed that **Model 6** appeared to be a robust choice for correcting the OPCs in a mobile setting.

This analysis represents one of the first attempts to provide guidance on correcting LCS for use in mobile deployments. We found that although more complex machine learning algorithms appeared to perform best at the stationary collocation site, simpler models appeared to transfer better to the mobile setting. This is likely because more complex models are overfit to the stationary site, even after appropriate cross-validation, due to differences in conditions such as meteorology or aerosol type at the stationary site compared to during the mobile deployment. Importantly, we also observed that corrections developed using high-resolution measurements at the stationary co-location site, that were similar to the frequency of data collection during the mobile deployment, also transferred best to the mobile setting. We note however, that the best correction model only yielded a low to moderate correlation (R = 0.29 when comparind the DustTrak with OPC 1 and 0.59 when comparing the DustTrak with OPC 2) between the corrected OPC and DustTrak data.

*Table 1: Comparing corrected OPC measurements using different correction models derived from minute- and hourly-averaged collocated data with that of the corresponding (from the correction model developed using 1 minute-averaged collocation data) DustTrak data in our mobile deployment. In Column Model: $s_1$-$s_{15}$ and b are empirically derived from regression analyses. Comparisons using OPC data corrected with Model 6 and the DustTrake data in the mobile setting are highlighted.*

| ID | Name | Model | Mobile deployment | | | | | | | | | | | |
|---|---|---|---|---|---|---|---|---|---|---|---|---|---|---|
| | | | Correction derived from minute-averaged OPC data | | | | | | Correction derived from hourly-averaged OPC data | | | | | |
| | | | DustTrak and OPC 1 | | DustTrak and OPC 2 | | OPC 1 and OPC2 | | DustTrak and OPC 1 | | DustTrak and OPC 2 | | OPC 1 and OPC 2 | |
| | | | R | RMSE | R | RMSE | R | RMSE | R | RMSE | R | RMSE | R | RMSE |
| | **Raw OPC measurements** | | | | | | | | | | | | | |
| 0 | Raw | | 0.26 | 6.3 | 0.20 | 6.6 | 0.41 | 1.7 | 0.26 | 6.3 | 0.20 | 6.6 | 0.41 | 1.7 |
| | **Multivariate Regression** | | | | | | | | | | | | | |
| 1 | Linear | $PM_{2.5, corrected} = PM_{2.5} \times s1 + b$ | 0.26 | 6.9 | 0.20 | 3.4 | 0.41 | 1.9 | 0.26 | 16.9 | 0.20 | 3.5 | 0.41 | 4.4 |
| 2 | +RH | $PM_{2.5, corrected} = PM_{2.5} \times s_1 + RH \times s_2 + b$ | 0.26 | 5.8 | 0.20 | 3.4 | 0.41 | 1.5 | 0.26 | 16.7 | 0.20 | 3.5 | 0.41 | 4.3 |
| 3 | +T | $PM_{2.5, corrected} = PM_{2.5} \times s_1 +$ | 0.26 | 7.2 | 0.31 | 3.3 | 0.46 | 1.9 | 0.26 | 17.7 | 0.31 | 3.4 | 0.46 | 4.6 |

| | | | | | | | | | | | | | |
|---|---|---|---|---|---|---|---|---|---|---|---|---|---|
| | | $T \times s_2 + b$ | | | | | | | | | | | |
| 4 | +D | $PM_{2.5, corrected} = PM_{2.5} \times s_1 + D \times s_2 + b$ | 0.28 | 5.0 | 0.42 | 3.7 | 0.51 | 1.2 | 0.27 | 12.6 | 0.39 | 3.8 | 0.48 | 3.3 |
| 5 | +RH x T | $PM_{2.5, corrected} = PM_{2.5} \times s_1 + RH \times s_2 + T \times s_3 + RH \times T \times s_4 + b$ | 0.27 | 5.3 | 0.33 | 3.5 | 0.46 | 1.3 | 0.26 | 14.9 | 0.31 | 3.6 | 0.45 | 3.9 |
| 6 | +RH x D | $PM_{2.5, corrected} = PM_{2.5} \times s_1 + RH \times s_2 + D \times s_3 + RH \times D \times s_4 + b$ | 0.29 | 5.6 | 0.59 | 3.9 | 0.65 | 1.4 | 0.27 | 15.6 | 0.56 | 4.0 | 0.55 | 4.1 |
| 7 | +D x T | $PM_{2.5, corrected} = PM_{2.5} \times s_1 + D \times s_2 + T \times s_3 + D \times T \times s_4 + b$ | 0.28 | 5.2 | 0.48 | 3.7 | 0.55 | 1.2 | 0.27 | 13.9 | 0.45 | 3.7 | 0.51 | 3.6 |
| 8 | +RH x T x D | $PM_{2.5, corrected} = PM_{2.5} \times s_1 + RH \times s_2 + T \times s_3 + D \times s_4 + RH \times T \times s_5 + RH \times D \times s_6 + T \times D \times s_7 + RH \times T \times D \times s_8 + b$ | 0.25 | 5.7 | 0.11 | 3.8 | 0.40 | 1.41 | 0.26 | 16.4 | 0.08 | 3.8 | 0.38 | 4.2 |
| 9 | PM x RH | $PM_{2.5, corrected} = PM_{2.5} \times s_1 + RH \times s_2 + RH \times PM_{2.5} \times s_3 + b$ | 0.28 | 5.4 | 0.31 | 3.6 | 0.41 | 1.7 | 0.26 | 14.8 | 0.26 | 3.9 | 0.39 | 4.0 |
| 10 | PM x D | $PM_{2.5, corrected} = PM_{2.5} \times s_1 + D \times s_2 + D \times PM_{2.5} \times s_3 + b$ | 0.17 | 18.4 | 0.28 | 4.6 | 0.53 | 6.7 | 0.19 | 36.5 | 0.27 | 5.1 | 0.53 | 13.0 |
| 11 | PM x T | $PM_{2.5, corrected} = PM_{2.5} \times s_1 + T \times s_2 + T \times PM_{2.5} \times s_3 + b$ | 0.25 | 8.1 | 0.26 | 3.6 | 0.39 | 2.6 | 0.26 | 18.0 | 0.24 | 3.9 | 0.38 | 4.9 |
| 12 | PM x nonlinear RH | $PM_{2.5, corrected} = PM_{2.5} \times s_1 + \frac{RH^2}{(1-RH)} \times s_2 + \frac{RH^2}{(1-RH)} \times PM_{2.5} \times s_3 + b$ | 0.22 | 19.2 | 0.20 | 4.4 | 0.42 | 6.0 | 0.23 | 40.4 | 0.20 | 5.0 | 0.42 | 11.9 |
| 13 | PM x RH x T | $PM_{2.5, corrected} = PM_{2.5} \times s_1 + RH \times s_2 + T \times s_3 + PM_{2.5} \times RH \times s_4 + PM_{2.5} \times T \times s_5 + RH \times T \times s_6 + PM_{2.5} \times RH \times T \times s_7 + b$ | 0.18 | 24.0 | 0.27 | 5.3 | 0.52 | 7.8 | 0.20 | 44.2 | 0.25 | 6.9 | 0.49 | 14.8 |
| 14 | PM x RH x D | $PM_{2.5, corrected} = PM_{2.5} \times s_1 + RH \times s_2 + D \times s_3 + PM_{2.5} \times RH \times s_4 + PM_{2.5} \times D \times s_5 + RH \times D \times s_6 + PM_{2.5} \times RH \times D \times s_7 + b$ | 0.21 | 17.3 | 0.28 | 4.5 | 0.41 | 5.9 | 0.21 | 42.3 | 0.23 | 6.9 | 0.45 | 14.8 |
| 15 | PM x T x D | $PM_{2.5, corrected} = PM_{2.5} \times s_1 + T \times s_2 + D \times s_3 + PM_{2.5} \times T \times s_4 + PM_{2.5} \times D \times s_5 + T \times D \times s_6 + PM_{2.5} \times T \times D \times s_7 + b$ | 0.17 | 43.6 | 0.29 | 5.7 | 0.55 | 9.8 | 0.19 | 56.0 | 0.27 | 7.9 | 0.54 | 16.8 |
| 16 | PM x RH x T x D | $PM_{2.5, corrected} = PM_{2.5} \times s_1 + RH \times s_2 + T \times s_3 + D \times s_4 + PM_{2.5} \times RH \times s_5 + PM_{2.5} \times T \times s_6 + T \times RH \times s_7 + PM_{2.5} \times D \times s_8 + D \times RH \times s_9 + D \times T \times s_{10} + PM_{2.5} \times$ | 0.20 | 33.1 | 0.17 | 5.4 | 0.47 | 5.3 | 0.11 | 101.2 | 0.19 | 8.8 | 0.60 | 39.3 |

| | | | | | | | | | | | | | |
|---|---|---|---|---|---|---|---|---|---|---|---|---|---|
| | | $RH \times T \times s_{11} + PM_{2.5} \times RH \times D \times s_{12} + PM_{2.5} \times D \times T \times s_{13} + D \times RH \times T \times s_{14} + PM_{2.5} \times RH \times T \times D \times s_{15} + b$ | | | | | | | | | | | |
| | **Machine Learning (10-fold CV)** | | | | | | | | | | | | |
| 17 | Random Forest | $PM_{2.5, corrected} = f(PM_{2.5}, T, RH)$ | 0.17 | 3.6 | 0.16 | 4.1 | 0.09 | 1.4 | 0.25 | 2.7 | 0.17 | 4.0 | 0.14 | 2.3 |

## 3.2 Evaluating factors that best explained differences between corrected OPC and DustTrak and OPC 1 and OPC 2 measurements

The results from our ANOVA analyses are displayed in **Table 2**. The bulk of the variation in the absolute difference/difference between the uncorrected OPC measurements remains unexplained. However, when evaluating factors that contributed to the variation in differences between corrected OPC measurements, we see T explains a moderate amount of variation (11.01% from 0.09% when looking at the absolute difference between corrected versus uncorrected OPC measurements; and 14.25 % from 1.11%, when looking at the difference between corrected versus uncorrected OPC measurements), followed by time-of-day (~3% from 0.2 - 0.3%). **Model 6** corrects OPC measurements using RH and D (but not explicitly T, although D and T are related). Thus, this result could arise from our choice of correction model (**Table 2**).

When evaluating factors that contribute to the differences/ absolute differences between corrected OPC and DustTrak measurements, we found that for OPC 1 >90% of the variation was unexplained by the factors considered. For OPC 2, T explained >20% of the variation (**Table 2**). This could be because **Model 6** did not adequately correct OPC 2 measurements for T as T during the collocation was not representative of T during the mobile deployment; whereas for OPC 1, as the mobile deployment occurred over more days, the T experienced was more representative (**Figure S3**). It is thus important to ensure that the environmental conditions during collocation are representative of conditions experienced during the mobile deployment. Another potential explanation is that degradation in the performance of OPC 1 due to aging was responsible for the largest fraction of the difference in OPC 1 and DustTrak measurements, rendering the impact of T small in comparison.

Date contributed next to the variation in the absolute difference/difference between corrected OPC 1 and DustTrak measurements, corresponding to 6.1% and 11.5%, respectively. For OPC 2, hour of day contributed marginally more to variation than day, accounting for ~4-5% of the variation in the difference and absolute difference of the corrected OPC 2 and DustTrak measurements (**Table 2**). Aerosol composition varies over the course of a day, and between days. Our results suggest that this variation does not greatly impact the difference between corrected-OPC and DustTrak results.

SVF, speed and orientation, even when they significantly contributed to variation of the differences between OPC and DustTrak measurements, or coincident OPC measurements, only accounted for < 1% of total variation. In supplementary analyses, we also considered road class and found that it also significantly contributed to a small proportion (< 5%) of total variation in these outcomes (**Table S4**).

Our analysis thus suggests, although vehicle speed, orientation of the OPC-N3, and sky view factor of the road the measurement was made on (a proxy for the street canyon effect) impacted the variation of the mobile corrected OPC-N3 measurements, their impact was small and do not greatly impact our mobile monitoring results. The large unexplained variation (> 80%) in the difference between the corrected measurements of the two OPCs is due to the difference in age of the two instruments. The performance of the older OPC which was in operation for several years prior to the experiment has likely degraded over time. More work is needed to understand how aging impacts the performance of LCS. The majority of variation (> 80%) in the difference between the corrected measurements from OPC 1 and the DustTrak is also unexplained, likely due to the same reason.

*Table 2*: Sum of Squares (Explained variation (%) in the difference and absolute difference between uncorrected and corrected OPC measurements, and corrected OPC and DustTrak $PM_{2.5}$ measurements) from an ANOVA analysis. The T and RH used in ANOVA analyses comparing the OPC measurements came from City Scanner corresponding to OPC 1. When comparing OPC 1 and OPC 2 measurements with DustTrak data, we used T and RH data corresponding to the City Scanner corresponding to OPC 1 and 2, respectively.

|  | OPC | No | Orientation | RH | T | Speed | Sky view factor | Day | Hour | Residuals | Total Sum of Squares |
|---|---|---|---|---|---|---|---|---|---|---|---|
| **Absolute difference between the uncorrected OPC $PM_{2.5}$ measurements** |  | 1 | - | 33* (0.17%) | 18* (0.09%) | 1 (0.01%) | 6 (0.03%) | 174* (0.90%) | 61* (0.31%) | 19,087 (98.49%) | 19,380 |
| **Absolute difference between the corrected OPC $PM_{2.5}$ measurements using Model 6 (derived from minute-averaged collocated measurements)** |  | 2 | - | 55* (0.51%) | 1,194* (11.01%) | 0 (0.00%) | 8* (0.07%) | 80* (0.74%) | 414* (3.82%) | 9,097 (83.86%) | 10,848 |
| **Absolute difference between the corrected DustTrak measurements and the OPC $PM_{2.5}$ measurements** | 1 | 3 | 318* (0.05%) | 1,252* (0.18%) | 206* (0.03%) | 429* (0.06%) | 1,171* (0.17%) | 41,850* (6.07%) | 5,672* (0.82%) | 638,062 (92.6%) | 688,960 |
|  | 2 | 4 | - | 129* (0.34%) | 15,247* (39.77%) | 45* (0.12%) | 4 (0.01%) | 1,476* (3.85%) | 1,739* (4.54%) | 19,697 (51.38%) | 38,337 |
| **Difference between the uncorrected OPC $PM_{2.5}$ measurements** |  | 5 | - | 16* (0.08%) | 228* (1.11%) | 1 (0.00%) | 14* (0.07%) | 261* (1.27%) | 40* (0.20%) | 19,924 (97.27%) | 20,484 |
| **Difference between the corrected OPC $PM_{2.5}$ measurements using Model 6 (derived from** |  | 6 | - | 29* (0.23%) | 1,828* (14.25%) | 3 (0.02%) | 26* (0.20%) | 234* (1.82%) | 337* (2.63%) | 10,370 (80.85%) | 12,827 |

| | | | | | | | | | | |
|---|---|---|---|---|---|---|---|---|---|---|
| minute-averaged collocated measurements) | | | | | | | | | | |
| Difference between the corrected DustTrak measurements and the OPC PM$_{2.5}$ measurements | 1 | 7 | 3,362* (0.33%) | 104,460* (10.20%) | 51,605* (5.04%) | 35 (0.00%) | 533* (0.05%) | 117,502* (11.47%) | 8,046* (0.79%) | 738,963 (72.13%) | 1,024,506 |
| | 2 | 8 | - | 122* (0.31%) | 15,192* (38.37%) | 50* (0.13%) | 1 (0.00%) | 1,352* (3.41%) | 1,640* (4.14%) | 21,239 (53.64%) | 39,596 |

*Note: The road class data we used had the following categories: 1) Motorways, 2) Major Roads, 3) Other Major Roads, 4) Secondary Roads, 5) Local Connecting Roads, 6) Local Roads of High Importance, 7) Local Roads, 8) Local Roads of Minor Importance, 9) Other Roads. SVF is a dimensionless parameter that indicates how much sky is obstructed by buildings. When the SVF is 0, the sky is totally obstructed. When the SVF is 1, there are no obstructions. The SVF for the sampling route was calculated elsewhere using Google Street View Imagery[28]*

## 3.3 Evaluating the use of mobile OPCs to detect hotspots

**Figure 1** displays hotspots identified using uncorrected OPC 1 and DustTrak measurements. We note that although there appear to be differences in the hotspots identified by each instrument, the most persistent hotspots corresponding to clusters that contain more than five measurements in the top percentile made in the same 100 m radius overlap. This result is remarkable despite concerns of degradation of measurements from OPC 1. This suggests that mobile OPCs can be used to identify important hotspots.

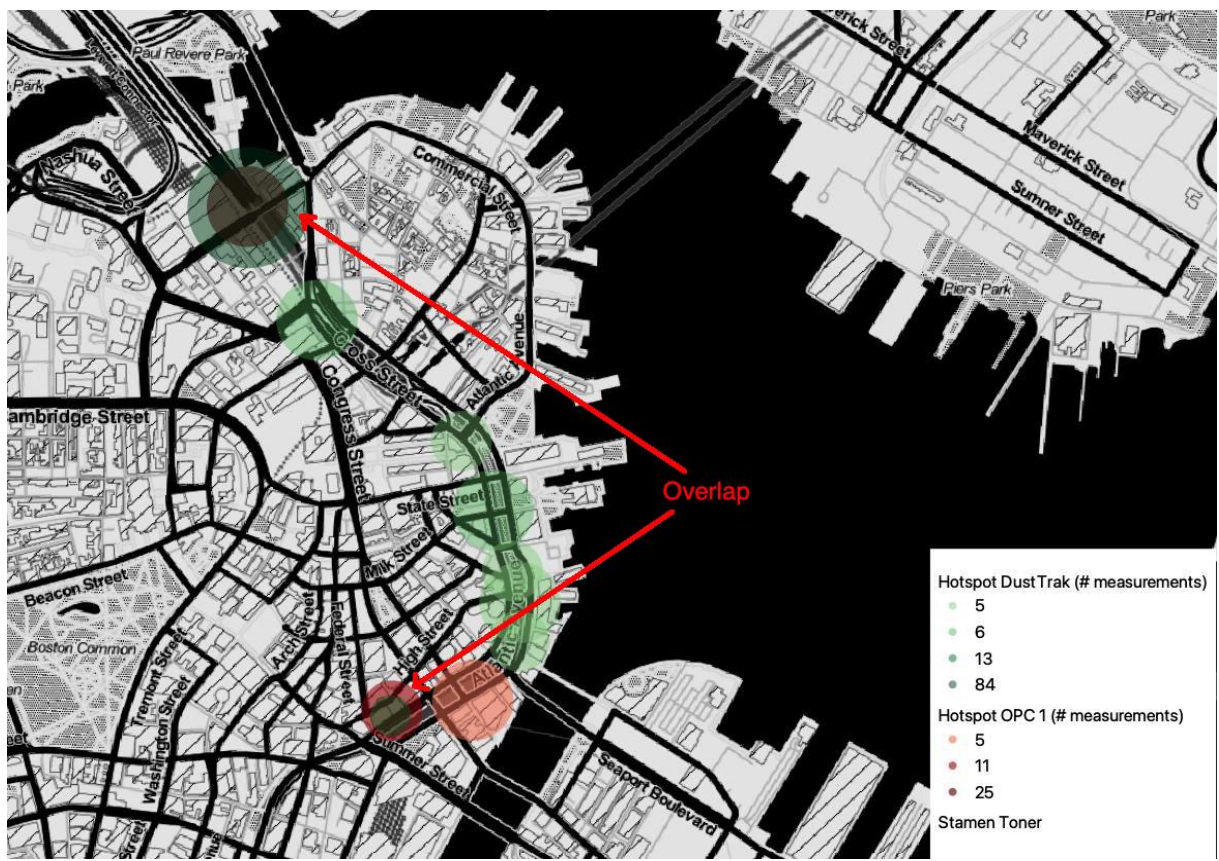

***Figure 1***: *Clusters of high PM$_{2.5}$ concentrations observed by OPC 1 and the DustTrak that include a minimum of 5 PM$_{2.5}$ observations in the top percentile recorded by each instrument. The legend displays the number of measurements in the top percentile corresponding to each cluster. The size of each cluster is proportional to the number of runs over which a high measurement was observed in the same 100 m radius.*

## 3.4 Evaluating the use of mobile OPCs to produce generalizable maps of PM$_{2.5}$ along the study route

**Figure S15** displays a 50 m × 50 m grid drawn across the study area. Grid cells in which mobile measurements were made are highlighted. **Figure 2** displays maps of generalizable PM$_{2.5}$ concentrations from mobile OPC 1, OPC 2 and DustTrak measurements. R and RMSE comparing generalizable OPC 1 concentrations and the corresponding DustTrak measurements were 0.39 and 4.9 μg/m$^3$, respectively. These metrics signify an improvement from comparing unaggregated corrected mobile OPC 1 and DustTrak measurements where the R and RMSE were 0.29 and 5.6, respectively (**Table 1**). R and RMSE comparing generalizable OPC 2 concentrations and the corresponding DustTrak measurements were 0.60 and 3.8 μg/m$^3$, respectively. These metrics also signified an improvement from comparing mobile OPC 1 and DustTrak measurements where the R and RMSE were 0.59 and 3.9 μg/m$^3$, respectively (**Table 1**).

The R and RMSE of generalizable OPC 1 and DustTrak measurements for only the 813 (out of 3,824) grid cells where the normalized standard error in the generalizable OPC 1 PM$_{2.5}$ concentrations calculated due to sampling variability from each instrument were < 20%, and measurements on the segment were made over more than single run improved to 0.38 and 2.8 µg/m³, respectively. The R and RMSE of generalizable OPC 2 and DustTrak measurements for only the 1,268 (out of 1,622) grid cells where the normalized standard error in the generalizable $PM_{2.5}$ concentrations calculated due to sampling variability from each instrument were < 20%, and measurements on the segment were made over more than single run were improved: 0.59 and 3.6 µg/m³, respectively (**Figure S16**). The reduced RMSE and similar R suggests that aggregating mobile LCS measurements to produce generalizable maps can reduce the error in these instruments. The error is smaller on segments with stable generalizable $PM_{2.5}$ concentrations.

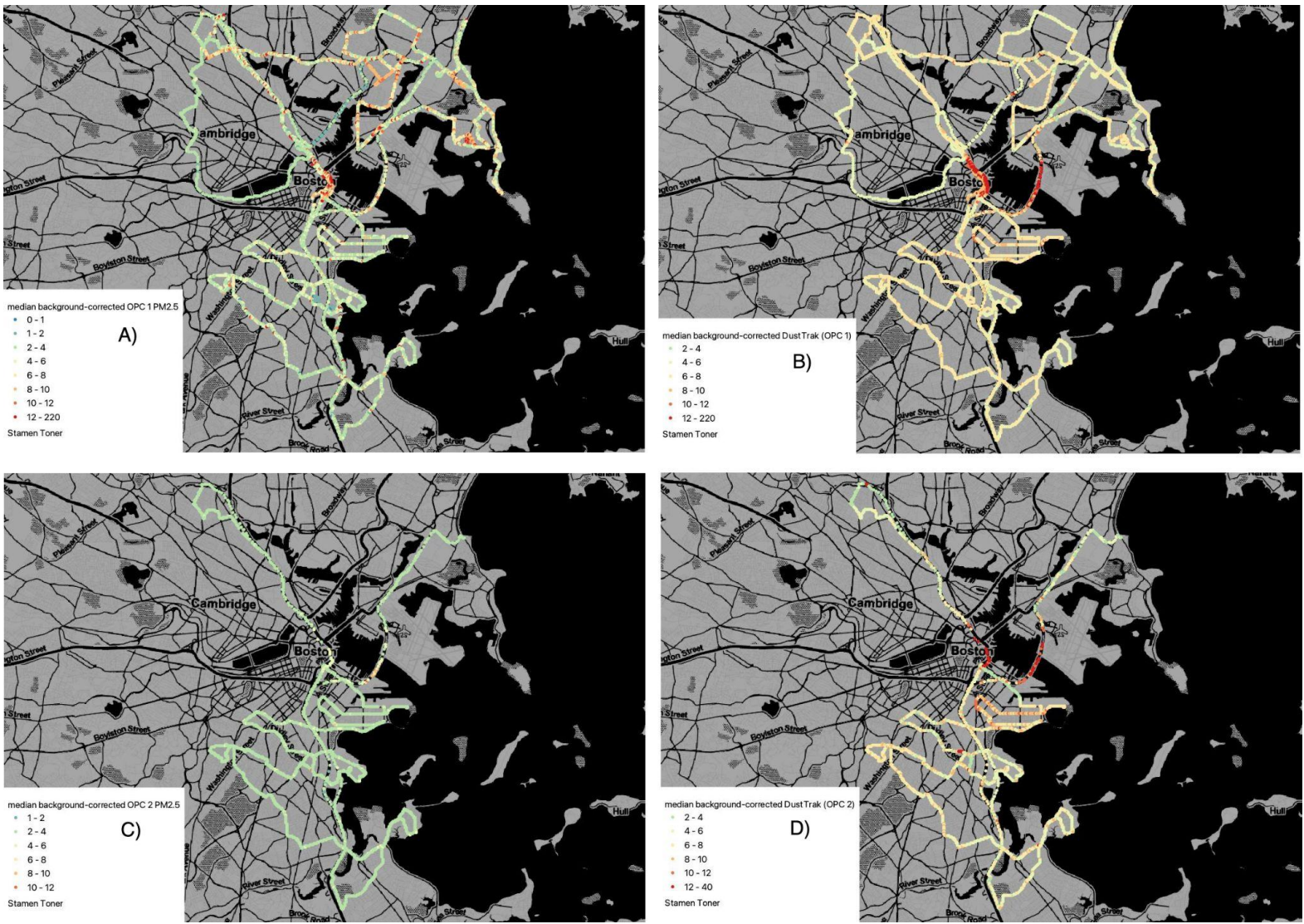

*Figure 2: Generalizable $PM_{2.5}$ concentrations (median background-corrected $PM_{2.5}$ concentrations displayed at the center of the corresponding 50 m × 50 m grid cell across the study area) produced using A) OPC 1 measurements, B) DustTrak measurements coincident with OPC 1, C) OPC 2 measurements, D) DustTrak measurements coincident with OPC 2.*

# 4 Limitations and Future Directions

This experiment had several limitations:

1) We performed the LCS testing in a single season, during daytime hours. The labor-intensive nature of this experiment precluded conducting long-term tests in more seasons and hours of the day. We hope to conduct longer-term experiments in the future.
2) We evaluated a single LCS: the Alphasense OPC-N3. In future experiments we aim to test a wider array of LCS. We also aim to evaluate LCS that measure other pollutants in addition to $PM_{2.5}$.
3) Although a diverse set of LCS installation scenarios was tested in this experiment, there are many configurations (orientation of the LCS relative to the direction of travel, vehicle speeds etc.) that a user might choose to implement that differ significantly from those tested in this evaluation protocol; thus the results of this experiment may have limited application to other configurations. In future experiments, we aim to evaluate other configurations over a broader range of experimental conditions.

This experiment represents one of the first attempts to provide guidance on calibrating LCS for use in mobile settings. In general, we recommend that researchers use simpler calibration models to correct mobile LCS data, as these models transfer better from stationary to mobile deployments. This paper also represents one of the first attempts to evaluate the feasibility of using mobile LCS data for two key-use cases: the identification of hotspots and the production of generalizable maps of $PM_{2.5}$ concentrations. We find that LCS can be used to identify persistent hotspots. We also find that LCS can be used to develop reasonably accurate generalizable maps of $PM_{2.5}$ in cities. Finally, by using OPCs of different ages, our paper represents one of the first efforts to show that the age of the LCS plays an important role in mobile experiments, and more work is needed to understand the impact of age on LCS performance.

# Supplementary Information

## S1: Performance of the OPC-N3

The Air Quality Sensor Performance Evaluation Center (AQ-SPEC) of the South Coast Air Quality Management District (SCAQMD) reported that in a laboratory evaluation[23] of three OPC-N3 sensors, 5-min average PM$_1$, PM$_{2.5}$ and PM$_{10}$ measurements were highly correlated ($R^2 > 0.99$) with measurements from a reference monitor, and the accuracy of the sensors was consistent over the range of PM$_1$ (accuracy was 11-14%), PM$_{2.5}$ (accuracy was 17-24%), and PM$_{10}$ (accuracy was 4-5%). In a field-setting, the correlations between the sensors and the reference monitor (5-min averages) were lower: the $R^2$ was ~ 0.80, 0.60, 0.50 for PM$_1$, PM$_{2.5}$, and PM$_{10}$, respectively.

## S2: Description of Random Forest

*Random forest (RF)*: RF is a decision-tree-based machine learning algorithm that has been shown to perform well in air quality predictions. Briefly, to generate a random forest model, the user specifies the maximum number of trees that make up the forest. Each tree is constructed using a bootstrapped random sample from the training data set. The origin node of the decision tree is split into sub-nodes by considering a random subset of the possible explanatory variables. Trees are split based on which of the explanatory variables in each subset is the strongest predictor of the outcome. This process of node splitting is repeated until a terminal node is reached[1]. For our random forest models, the terminal node was specified using a minimum node size of five data points per node.

## S3 10-fold Cross-Validation

Our 10-fold cross-validation (CV) exercise occurred in the following manner: The model fitting dataset was randomly split into 10 groups, with each group containing about 10% of the grid cells. In each cross-validation iteration, we select nine groups of the grid cells. All of the data from these grid cells are used to fit the model and make predictions of the remaining group. We repeated this process 10 times, until every group was predicted.

The CV process was used to tune two major RF hyperparameters: the number of decision trees ($n_{tree}$) and the number of predictors randomly tried at each split ($m_{try}$). Briefly, the optimal $m_{try}$ was determined by optimizing the "pseudo $R^2$": the fraction of variance of the PM$_{2.5c}$ measurements explained by the "out of bag" (OOB) predictions. An OOB prediction is based only on trees in the random-forest ensemble that were not trained using the PM$_{2.5c}$ observations being predicted. Each $m_{try}$ value was used to train 10 random-forests. The average pseudo $R^2$ was used to select the best $m_{try}$ value.

# S4: Calibrating the DustTrak

We compared the results from using three widely-used calibration methods, including the simple linear regression model in Equation (1), a widely-used the RH-adjusted empirical equation proposed by Laulainen (1993)[2] in Equation (2), and the RH-adjusted linear regressio model in Equation (3).

$$PM_{2.5, reference} = s_1 + s_2 \times PM_{2.5, DustTrak} \quad \ldots\ldots..(1)$$

$s_1$, and $s_2$ are empirically determined parameters determine from a regression analysis. $PM_{2.5, DustTrak}$ and $PM_{2.5, reference}$ are the coincident 1-minute averaged $PM_{2.5}$ concentrations reported by the DustTrak and reference site respectively. The R and RMSE obtained were 0.81 and 2.0 µg/m³, respectively. In supplementary analyses, we also re-ran these regressions using hourly-averaged data. The R and RMSE obtained were 0.77 and 3.3 µg/m³, respectively.

$$OR = s_1 + s_2 \times \frac{(RH/100)^2}{(1 - RH/100)} \quad \ldots\ldots (2)$$

In Equation (2) OR is the overestimation ratio of the DustTrak $PM_{2.5}$ relative to the reference monitor and is $\frac{PM_{2.5, DustTrak}}{PM_{2.5, reference}}$. We used RH measurements from the reference station for this correction. $s_1$ and $s_2$ are empirically determined parameters from a regression analysis. Many studies have reported $s_1 = 1$ and $s_2 = 0.25$. Using minute-averaged data, we obtained an $s_1 = 1$, and an $s_2 = 0.13$, and an R and RMSE corresponding to 0.81 and 6.8 µg/m³, respectively. In supplementary analyses, we also re-ran these regressions using hourly-averaged data, where obtained an R and RMSE of 0.78 and 7.8 µg/m³, respectively.

$$OR = s_1 + s_2 \times RH \quad \ldots\ldots (3)$$

The R and RMSE obtained from Equation 3 were 0.79 and 7.7 µg/m³, while that obtained from using hourly-averaged data were 0.80 and 6.6 µg/m³, respectively.

# S5 Developing maps of generalizable $PM_{2.5}$ concentrations for the time-period of the mobile deployment in the sampling area

## S5.1 Background Correction

We estimated background contribution for the uncorrected and corrected data for each OPC and the DustTrak using two different methods:

1) We adopted the lowest 10th percentile of the $PM_{2.5}$ concentrations measured by the instrument under consideration for a given hour during the run as the fixed background value for that run.
2) We used a time-series, spline-of-minimums approach, presented by Brantley et al. (2013), to estimate background $PM_{2.5}$ for each sampling run. We did this by: (a)

Applying a rolling 30-second mean to smooth the measurements, (b) Dividing the time series into discrete 10-minute segments and locating the minimum concentration in each segment, and (c) Fitting a smooth, thin-plate regression spline through the minimum concentrations.

We made the assumption that the background is temporally varying but spatially uniform.

After estimating the background $PM_{2.5}$ for each instrument, we performed a background time-of-day correction or standardization using **Equations 4** and **5** to account for the period during which the vehicle operated:

$$PM_{2.5c,i} = PM_{2.5,OPC\ i} - PM_{2.5,bkg,i} + PM_{2.5,bkg,median} \quad \ldots\ldots\ldots\ldots\ldots\ldots\ldots\ldots\ldots\ldots (4)$$

where $PM_{2.5,OPC,i}$ is the calibrated measurement for event i, $PM_{2.5,bkg,i}$ is the contemporaneous background value of pollution over the entire region, and $PM_{2.5,bkg,median}$ is the median of the $PM_{2.5,bkg}$ values for the sampling run. By subtracting the time-of-day-run-resolved regional background from the pollution measurement, we can now compare background-corrected/local $PM_{2.5}$ concentrations measured during different runs over space.

When the background $PM_{2.5}$ measurement value was estimated to be more than that of the concentration measured, we applied a multiplicative background-correction factor as displayed in **Equation 5**.

$$PM_{2.5c,i} = PM_{2.5,OPC,\ i} \times PM_{2.5,bkg,median} / PM_{2.5,bkg,i} \ldots\ldots\ldots\ldots\ldots\ldots\ldots\ldots\ldots\ldots (5)$$

Unlike **Equation 4**, this assures non-negative $PM_{2.5c,i}$ values, based on considering the background value assessed during the measurement hour as a fraction of the day's median background, thus reducing the contribution of possible measurement error due to treating the hourly value as an absolute quantity.

We found that both methods to calculate the background-corrected $PM_{2.5}$ for each instrument produced similar background-corrected values. The median differences between the two background-corrected concentrations for the uncorrected and corrected OPC 1, OPC 2 and the DustTrak measurements were < 5 %. **Figure S14** displays the background-corrected OPC 1, OPC 2, and DustTrak data during the mobile deployment and confirms that both methods to derive background $PM_{2.5}$ concentrations yield similar results. The median differences between the raw and each background-corrected concentration for the OPC 1, OPC 2 and the DustTrak were between 0 - 5%. Given the minimal differences in median $PM_{2.5c}$ from the different methods, we chose the splines-of-minimum approach to obtain background concentrations for both pollutants. This is supported by previous research that found this approach to be an effective way to account for background concentrations for a range of pollutants over a variety of meteorological conditions and sampling routes (Brantley et al., 2013).

## S5.2 Calculating the generalizable $PM_{2.5}$ concentrations across the sampling route

We divided the sampling area into grid cells of 50 m in length, using the 'Create Grid' QGIS Tool. We aggregate the background-corrected $PM_{2.5}$ concentrations made over different runs and derive the sampling error using the method proposed in deSouza et al., (2020).

First, we snapped each background-corrected $PM_{2.5}$ measurement to the grid cell that it fell in. This allows measurements made in the same grid cell to be analyzed as a group. Second, we selected the median (which is outlier resistant) of all background-corrected $PM_{2.5}$ measurements made on a given road segment as the 'generalizable $PM_{2.5}$ concentration' for that segment. Third, we used a set of bootstrap resampling procedures to quantify the effect of sample-to-sample variability and of sampling error on the median concentration for each road segment. As a metric of precision, we used the ratio of the standard error of median concentration to the median concentration itself.

# Figures

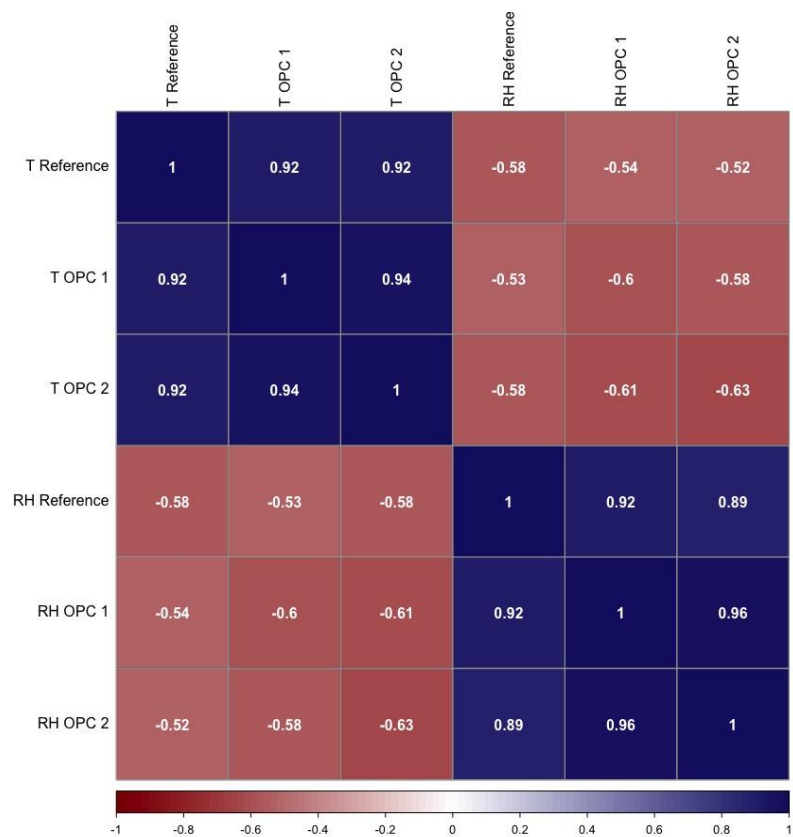

***Figure S1****: Pair-wise correlations between T and RH recorded by OPC 1, OPC 2, and by the reference monitoring station*

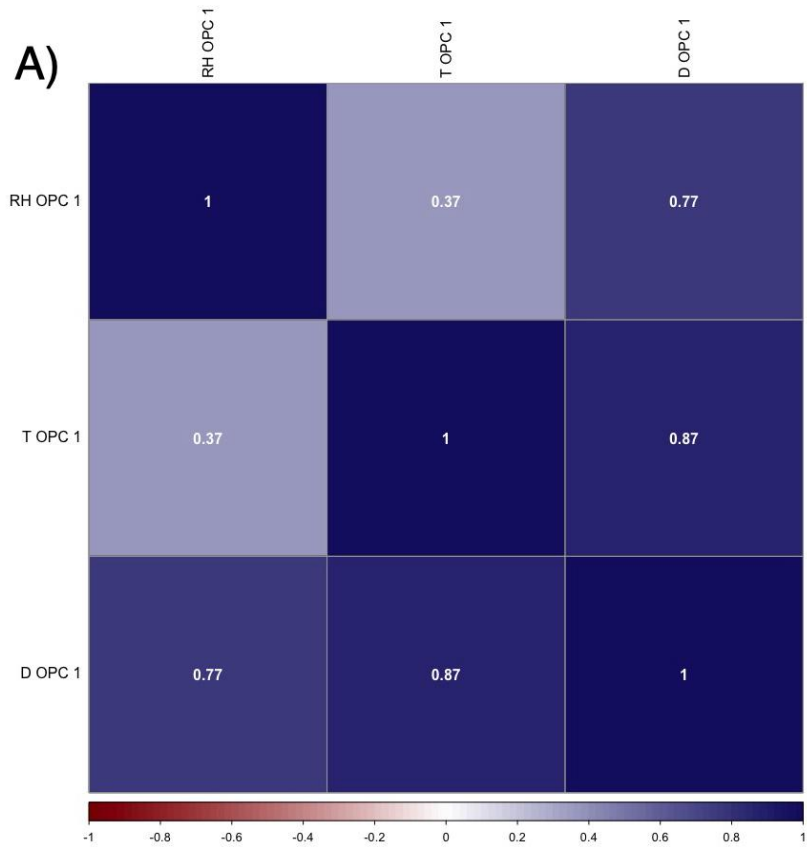
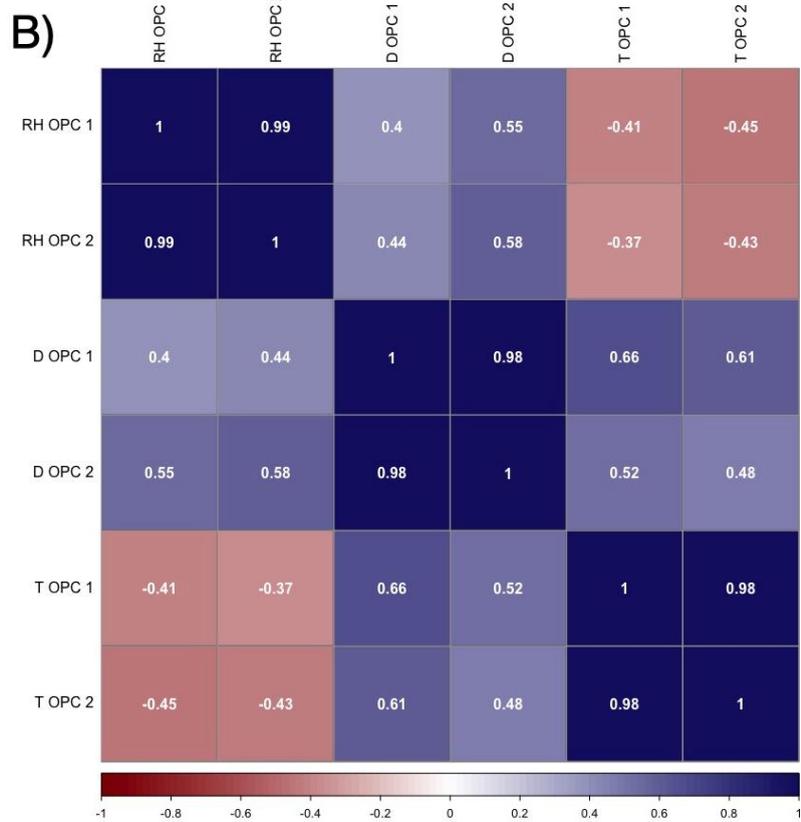

*Figure S2*: Pairwise correlations between T, RH and D for A) All mobile measurements made by the CityScanner platform corresponding to OPC 1, B) All coincident mobile

*measurements made by both CityScanner platforms for the time period where both OPC1 and OPC2 were operational*

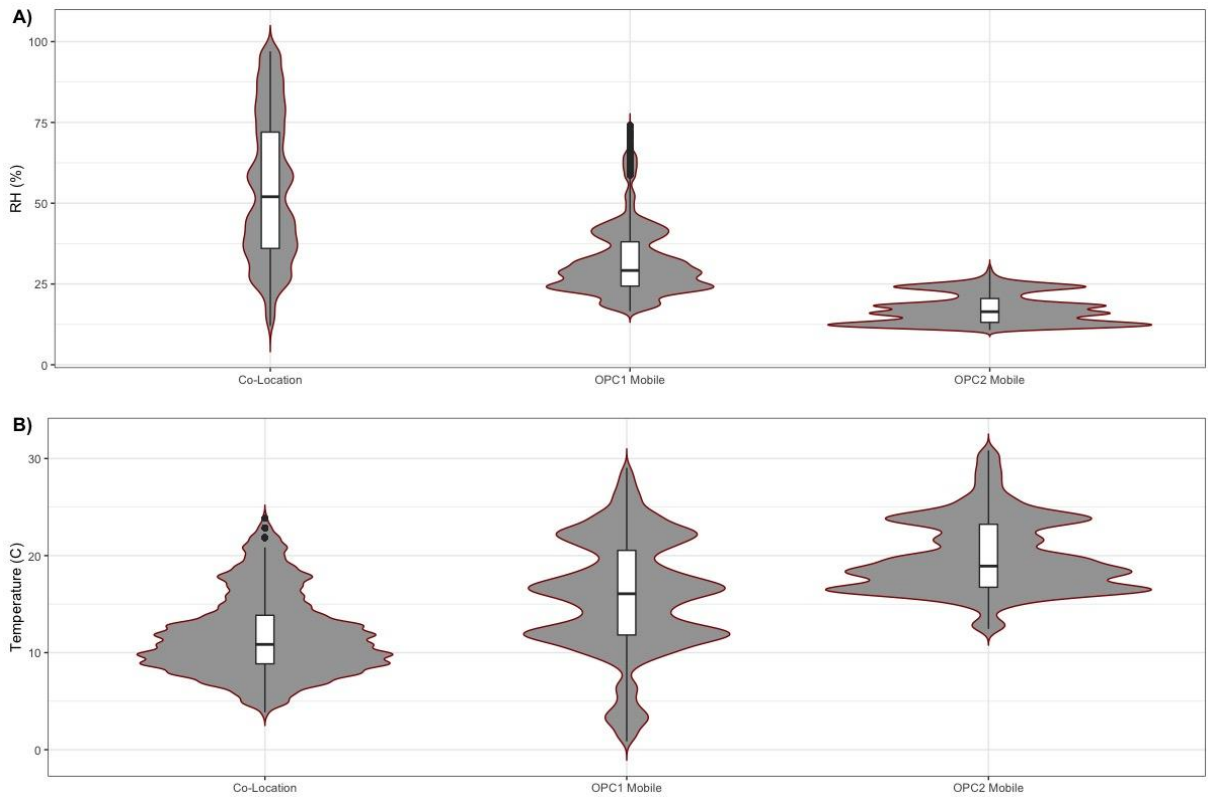

***Figure S3:*** *Comparing the distribution of Minute-level A) RH and B) T during the collocation of the OPCs in comparison to during the mobile monitoring deployment using violin and boxplots*

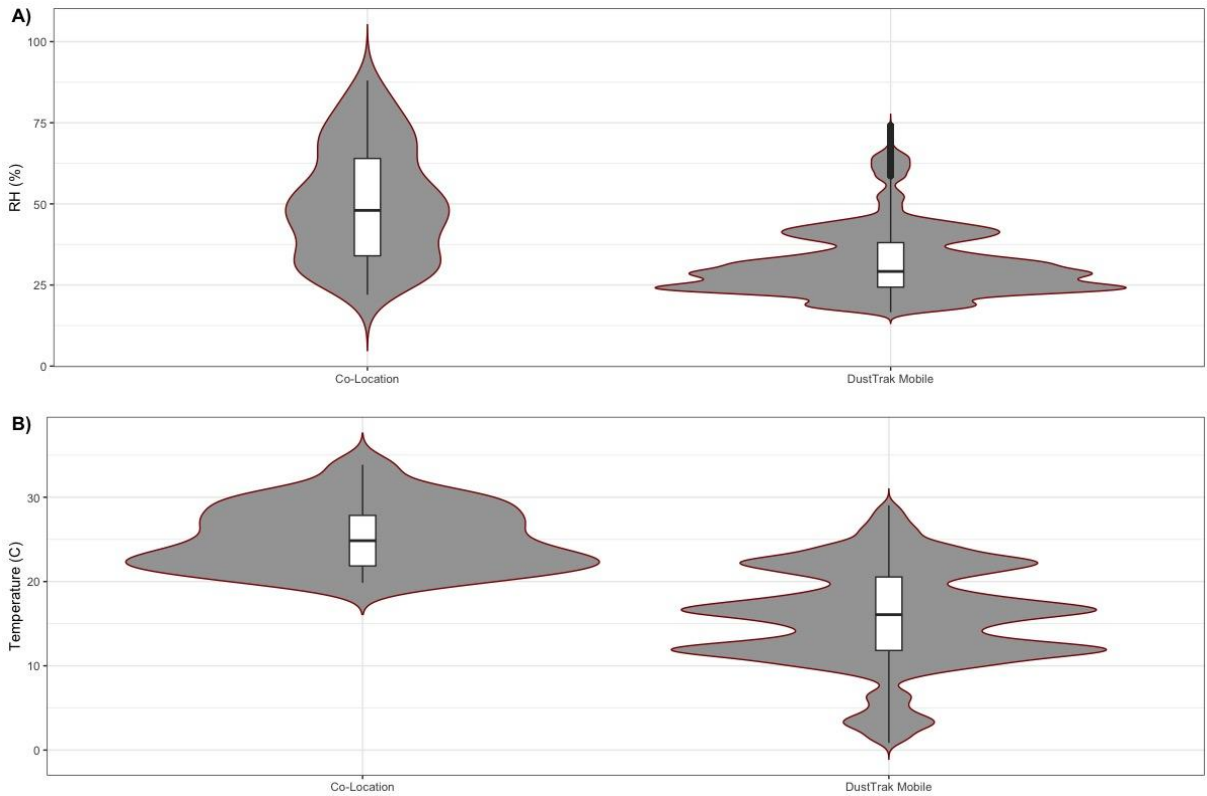

*Figure S4:* *Comparing the distribution of Minute-level A) RH and B) T during the collocation of the DustTrak in comparison to during the mobile monitoring deployment using violin and boxplots*

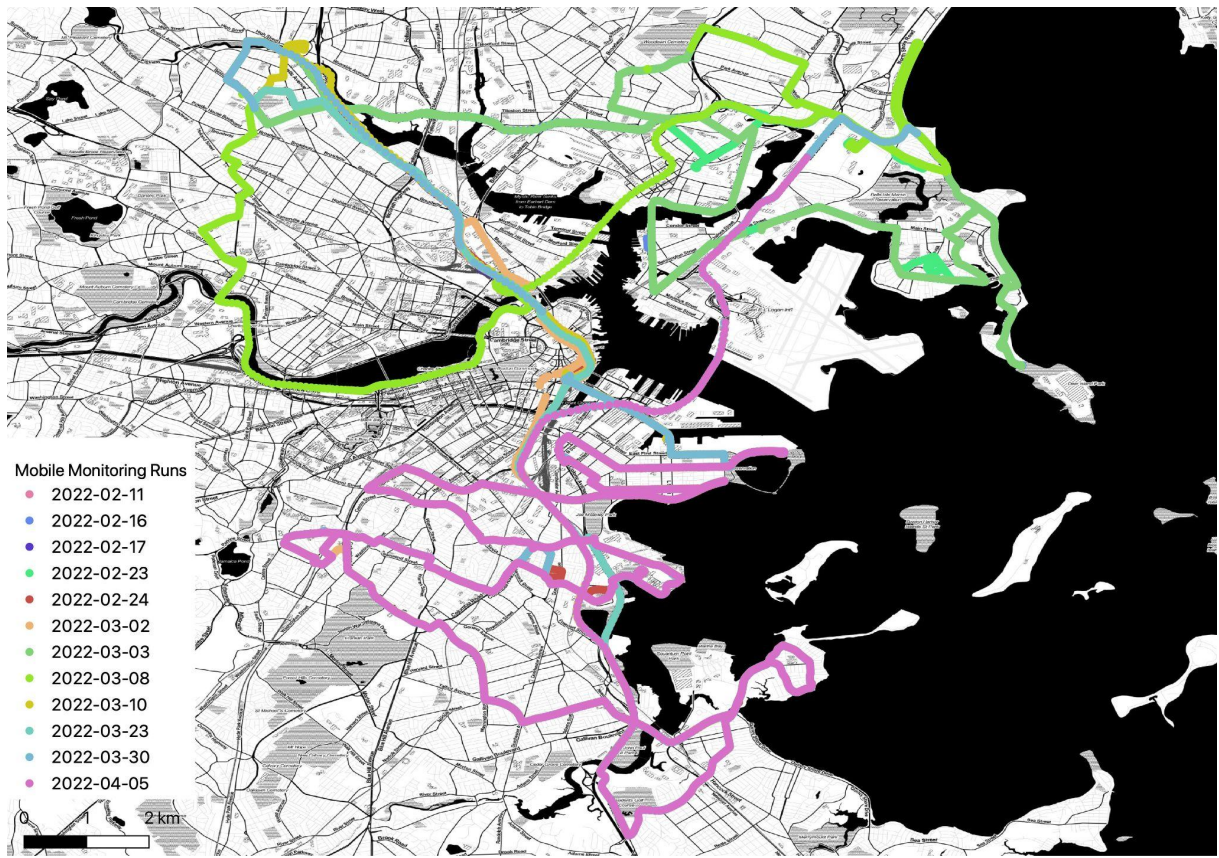

*Figure S5*: *Spatial extent of each of the 12 mobile monitoring runs*

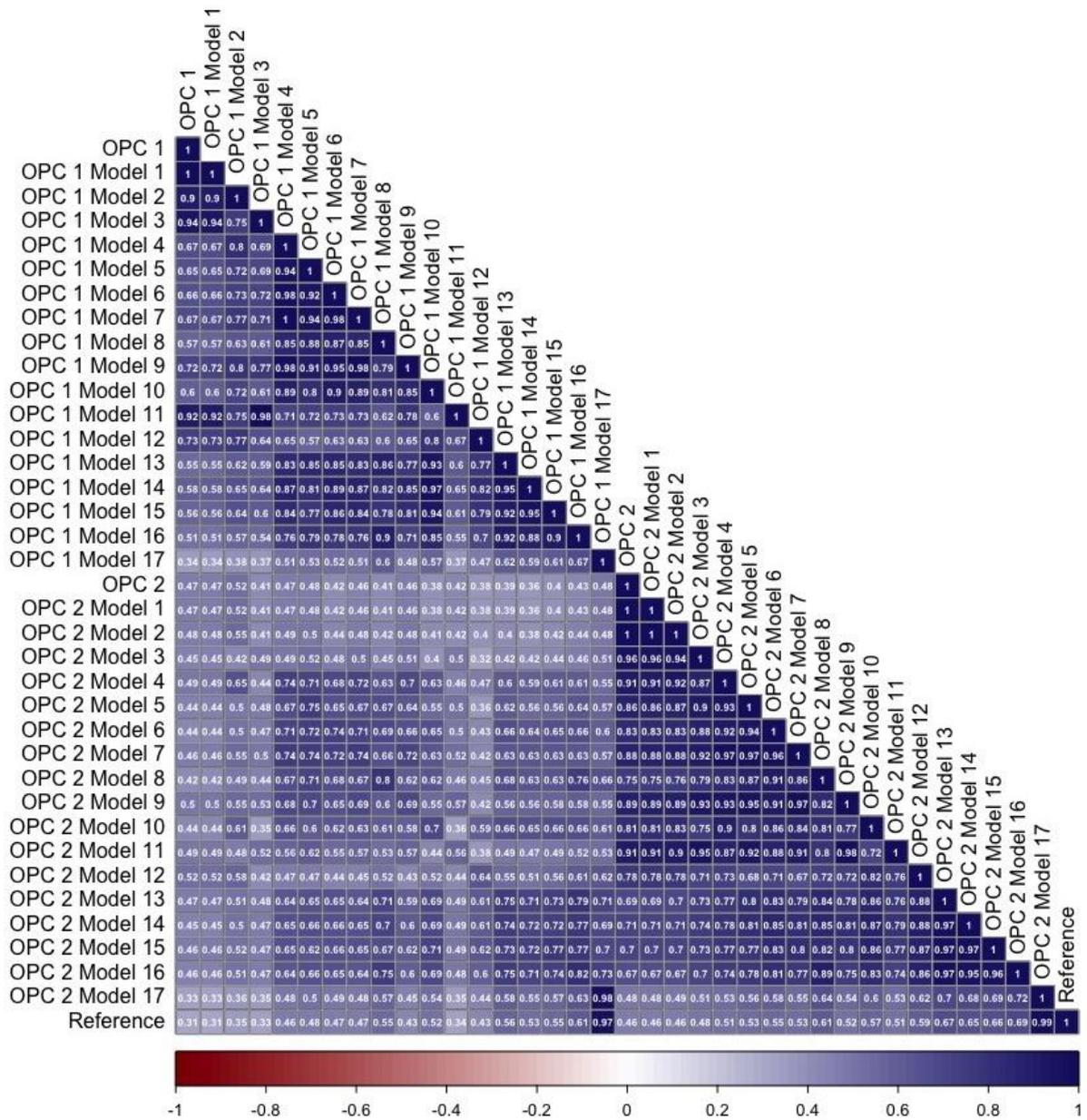

*Figure S6*: Pairwise correlations between OPC 1, OPC 2, corrected OPC 1 and OPC 2, and reference measurements during the stationary collocation experiment

*Figure S7*: *Pairwise correlations between OPC 1, OPC 2, corrected OPC 1 and OPC 2, and corrected DustTrak measurements during the mobile deployment*

*Figure S8*: Pairwise correlations between OPC 1, OPC 2, corrected (using hourly-averaged collocated measurements) OPC 1 and OPC 2, and corrected DustTrak measurements during the mobile deployment

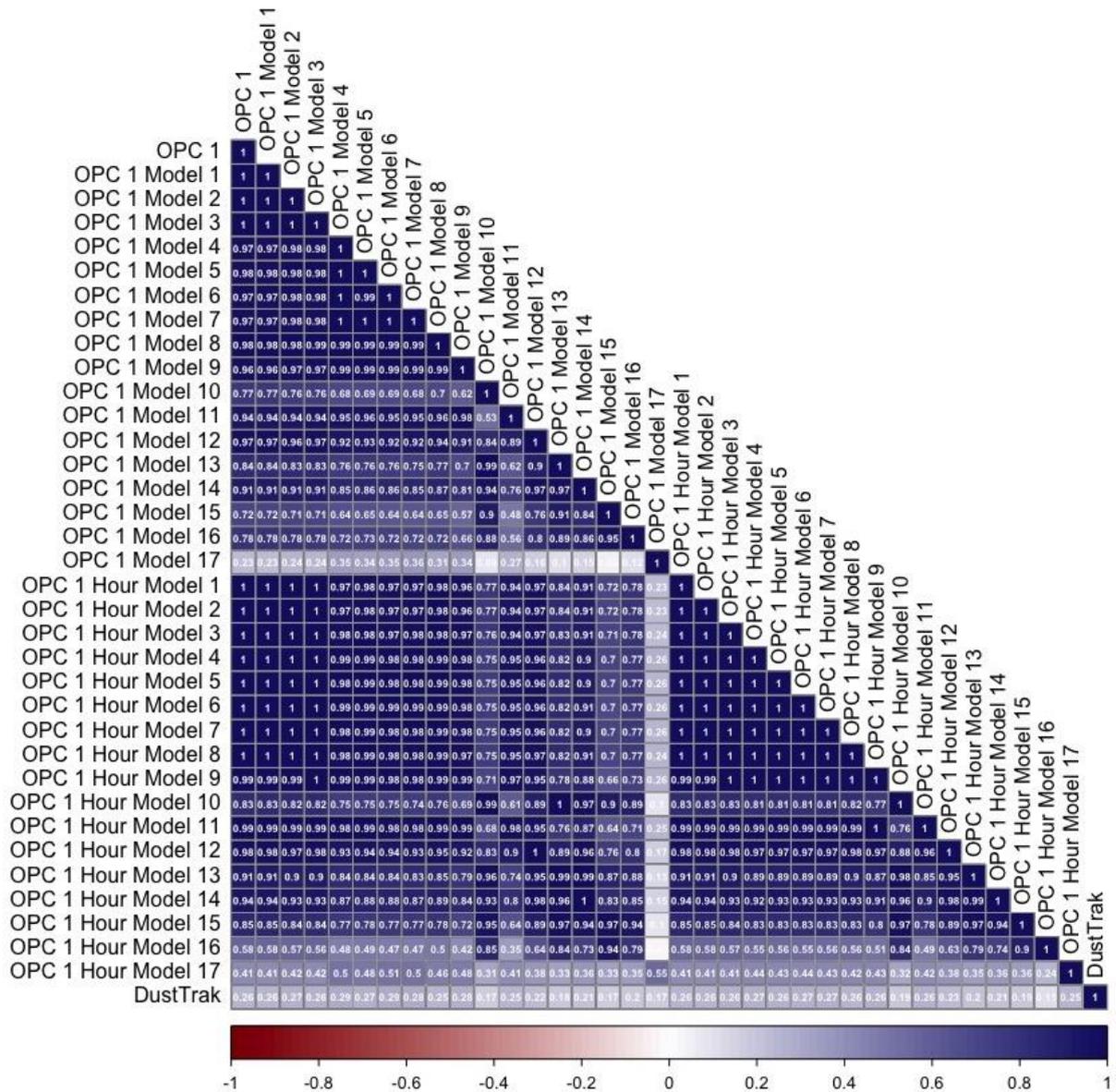

*Figure S9*: Pairwise correlations between OPC 1, corrected (using minute-averaged and hourly-averaged collocated measurements) OPC 1, and corrected DustTrak measurements during the mobile deployment

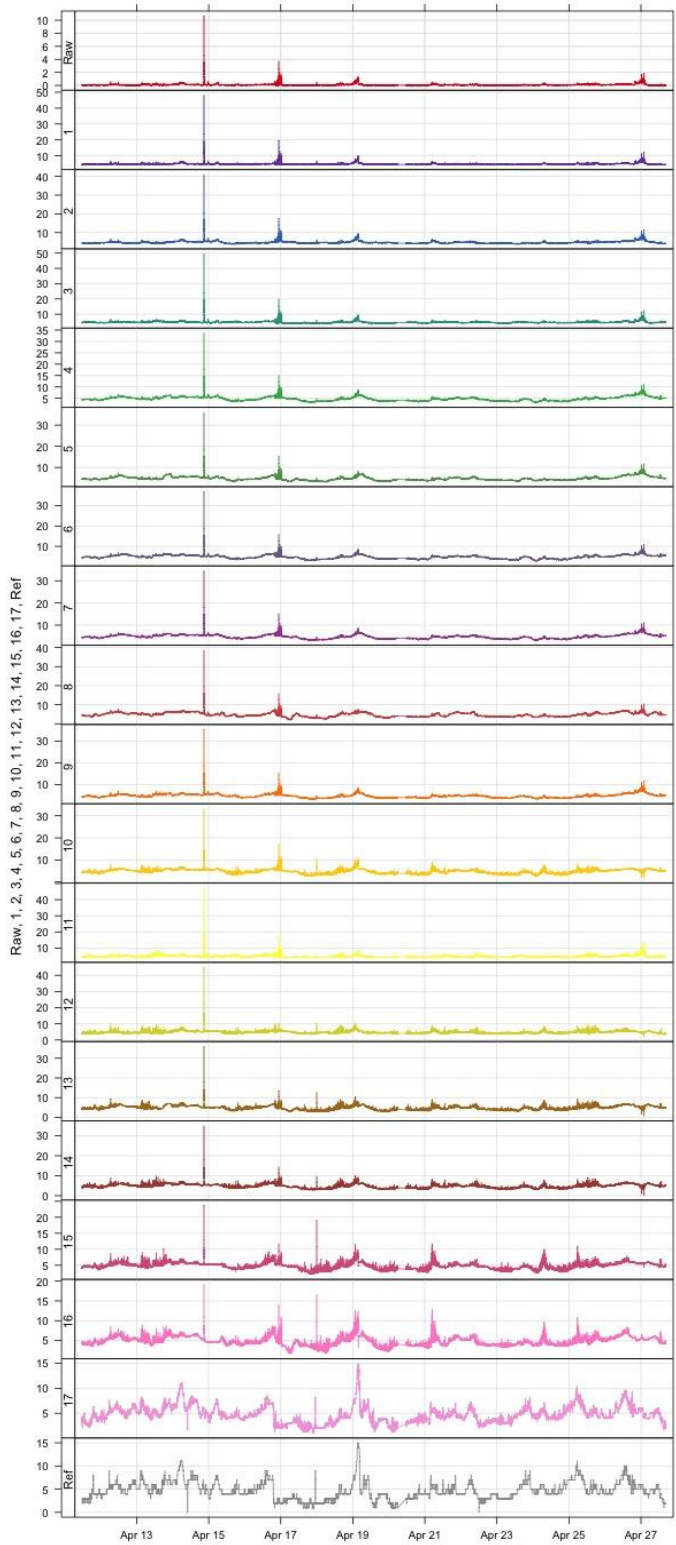

**Figure S10**: Reference, OPC 1, corrected OPC 1 measurements using the different calibration models (1 -17) described in Table 1 at the minute-averaged time resolution.

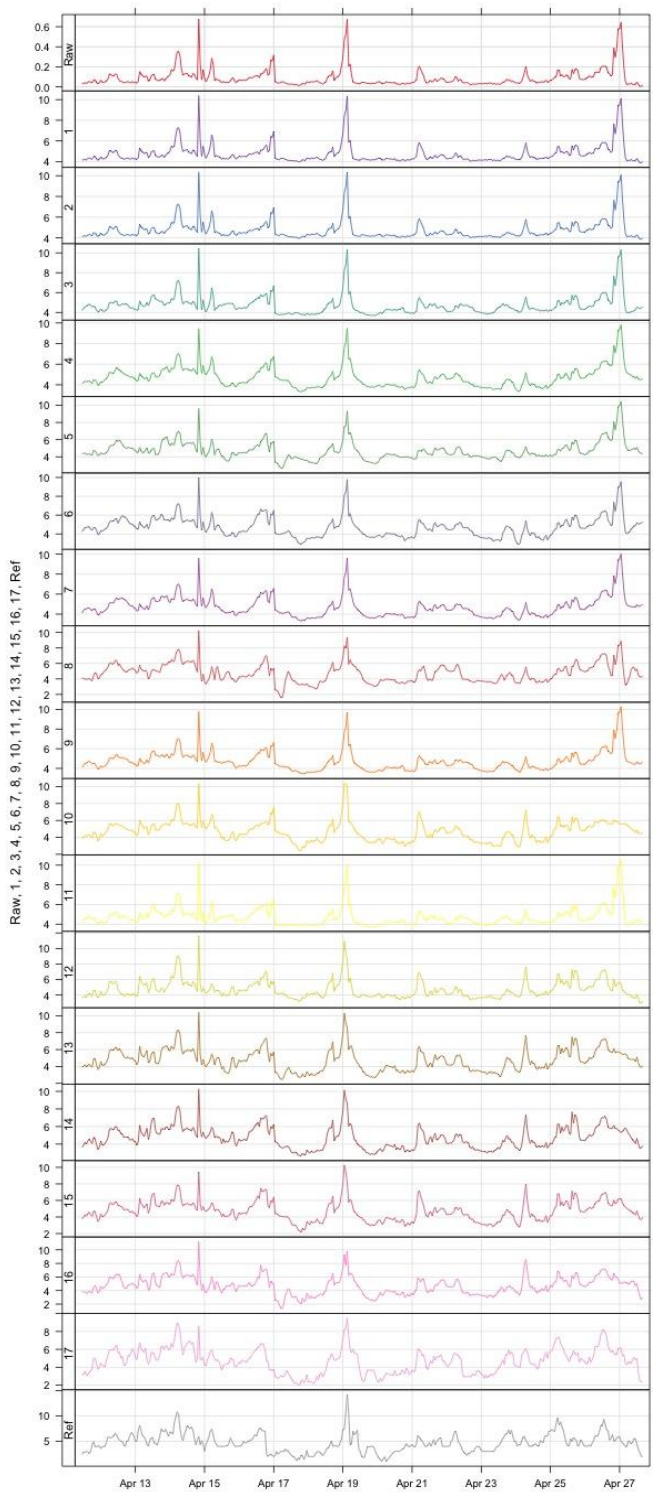

**Figure S11**: Reference, OPC 1, corrected OPC 1 measurements using the different calibration models (1 -17) described in Table 1 at the hourly-averaged time resolution.

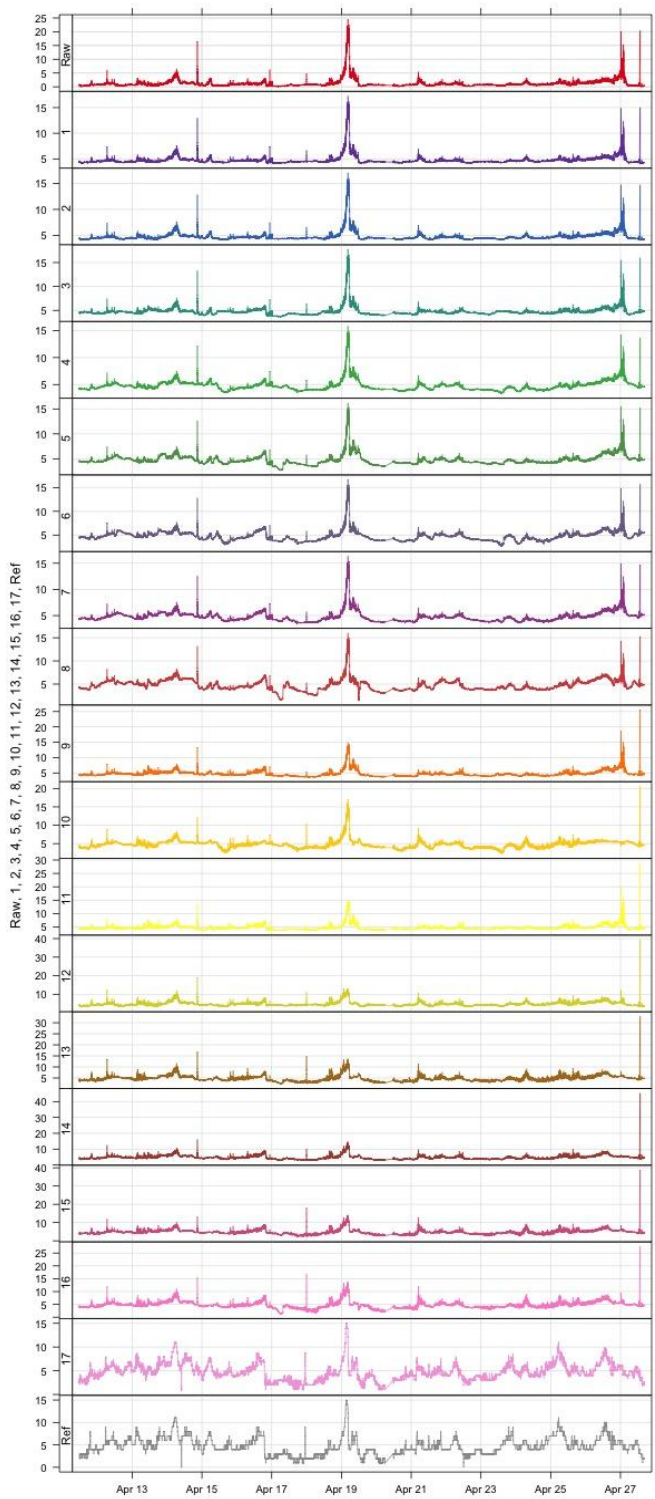

**Figure S12**: Reference, OPC 2, corrected OPC 2 measurements using the different calibration models (1 -17) described in Table 1 at the minute-averaged time resolution.

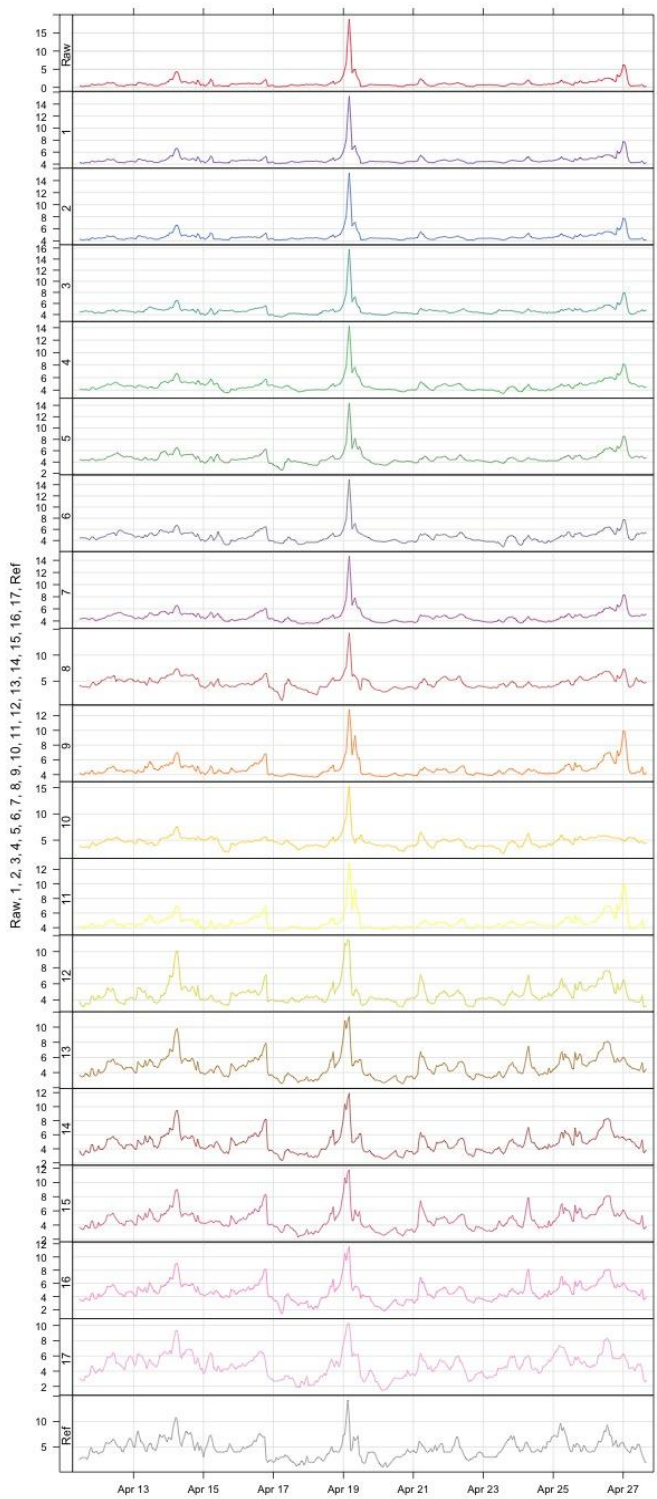

**Figure S13**: Reference, OPC 2, corrected OPC 2 measurements using the different calibration models (1 -17) described in Table 1 at the hourly-averaged time resolution.

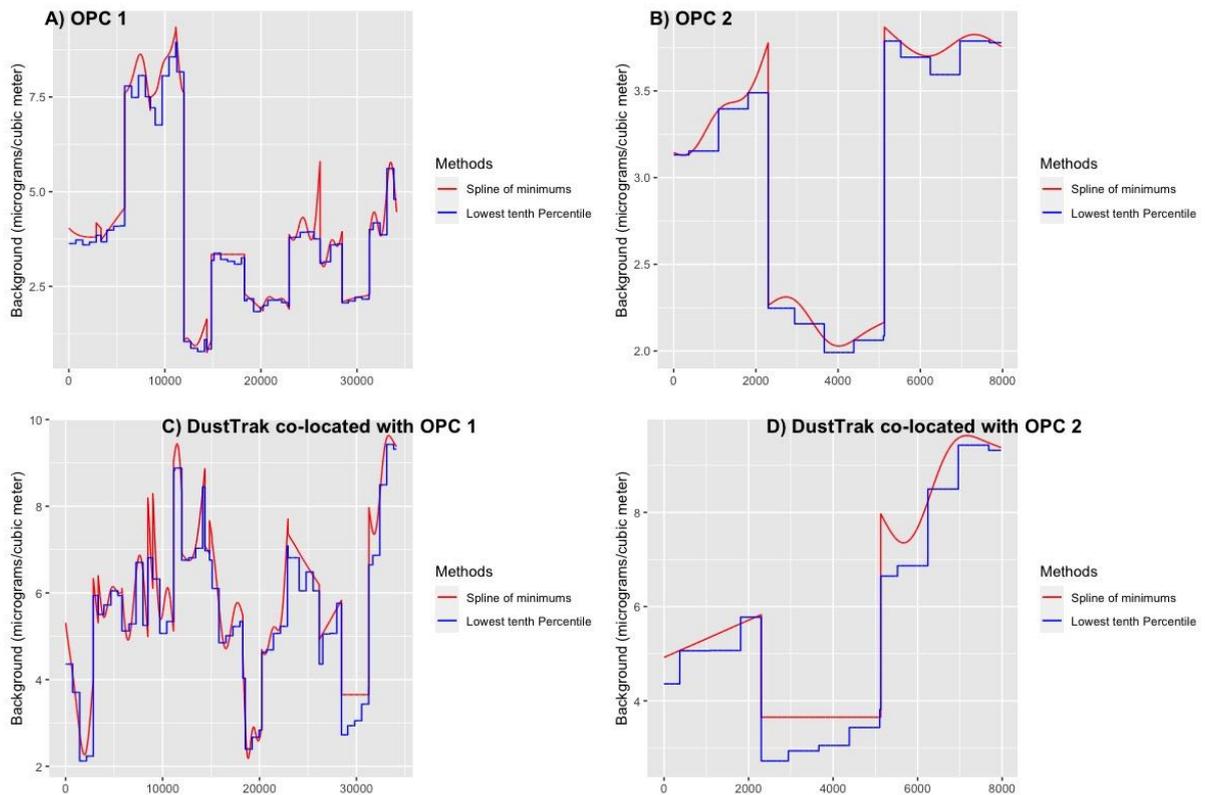

*Figure S14*: Comparing background PM$_{2.5}$ concentrations calculated for each instrument during the course of the mobile deployment

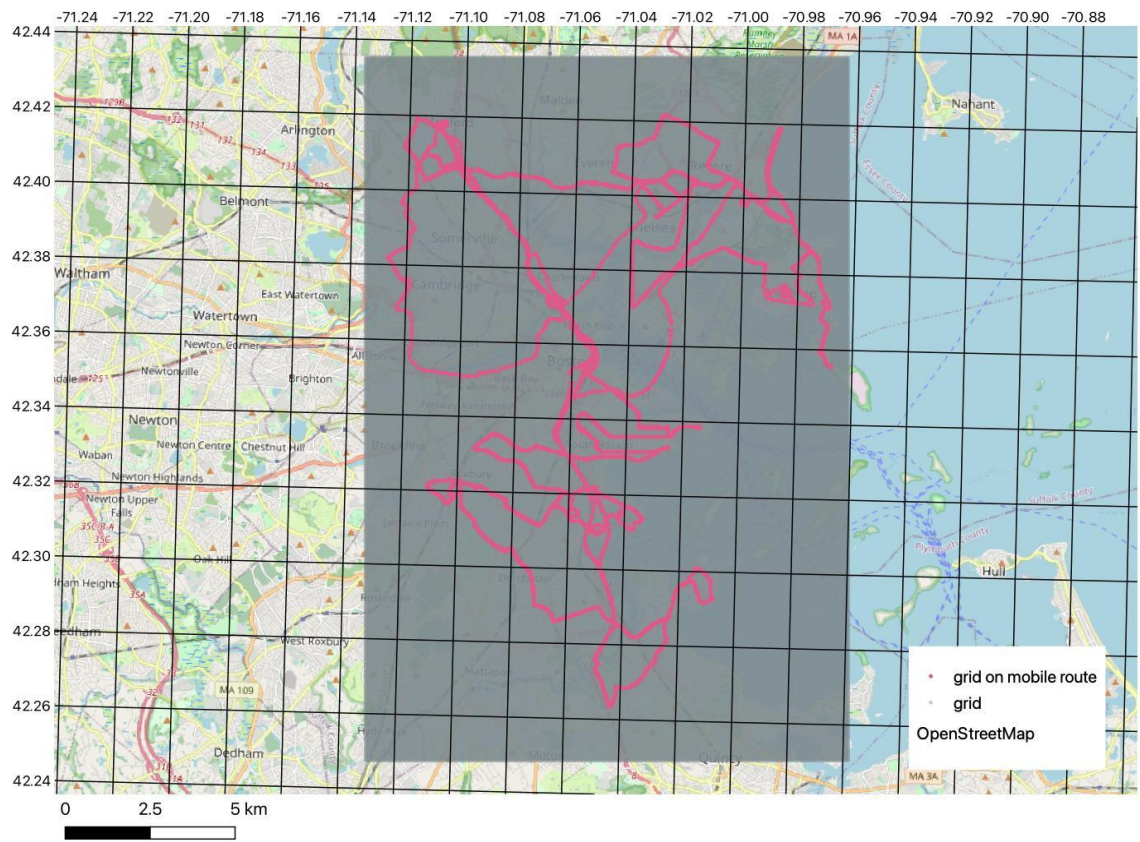

*Figure S15*: 50 m x 50 m grid cells drawn over the sampling area. The grid cells in which mobile PM$_{2.5}$ measurements were made are highlighted.

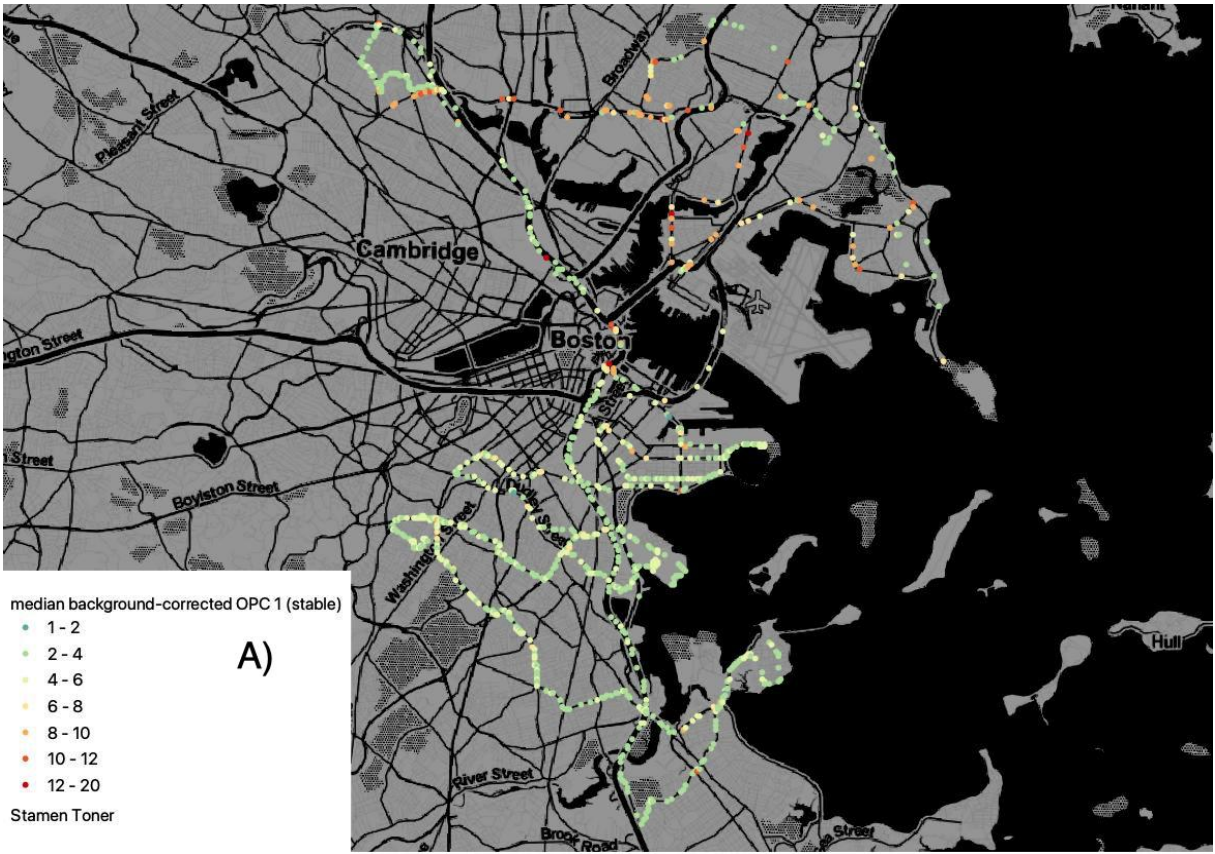

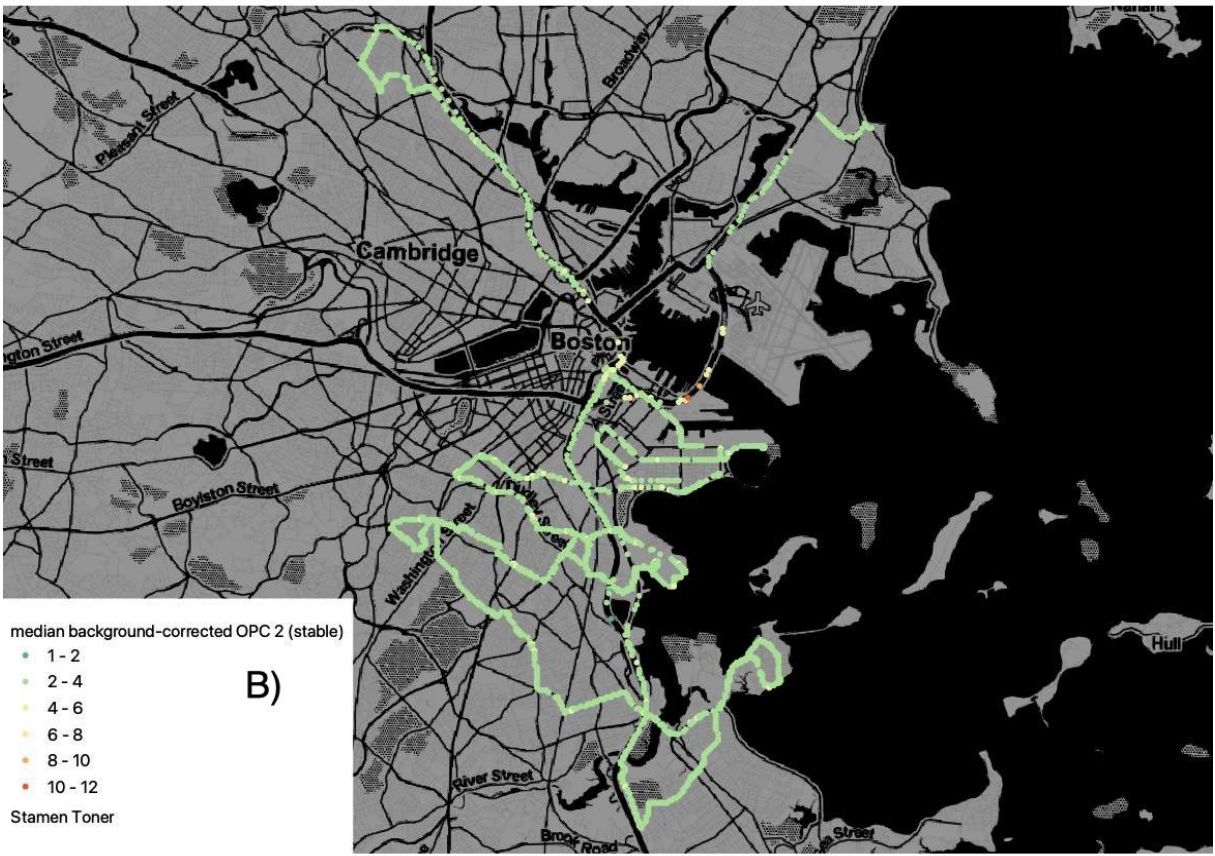

*Figure S16*: Generalizable PM$_{2.5}$ concentrations (median background-corrected calibrated PM$_{2.5}$ concentrations in 50 m x 50 m grid cells across the study area) produced using A) OPC 1 measurements, B) OPC 2 measurements for grid cells where the normalized error in the median PM$_{2.5}$ concentration was < 20% and measurements in that grid cell were made over > 1 day

# Tables

*Table S1*: Performance of the correction models using hourly-averaged and minute-averaged data as captured using root mean square error (RMSE), and Pearson correlation (R). A 10-fold CV was used to prevent overfitting in the machine learning models. Separate correction models were developed for the two OPCs used in this study. In Column Model: $s_1$-$s_{15}$ and b are empirically derived from regression analyses.

| ID | Name | Model | Minute-averaged | | | | Hourly-averaged | | | |
|----|------|-------|---|---|---|---|---|---|---|---|
|    |      |       | OPC 1 | | OPC 2 | | OPC 1 | | OPC 2 | |
|    |      |       | R | RMSE (µg/m³) | R | RMSE (µg/m³) | R | RMSE (µg/m³) | R | RMSE (µg/m³) |
|    | **Raw OPC measurements** | | | | | | | | | |
| 0  | Raw  |       | 0.31 | 5.0 | 0.46 | 4.0 | 0.50 | 5.0 | 0.51 | 3.9 |
|    | **Multivariate Regression** | | | | | | | | | |
| 1  | Linear | PM$_{2.5,\,corrected}$ = PM$_{2.5}$ × s1 + b | 0.31 | 1.8 | 0.46 | 1.7 | 0.50 | 1.6 | 0.51 | 1.6 |
| 2  | +RH | PM$_{2.5,\,corrected}$ = PM$_{2.5}$ × s$_1$ + RH × s$_2$ + b | 0.35 | 1.8 | 0.46 | 1.7 | 0.50 | 1.6 | 0.51 | 1.6 |
| 3  | +T | PM$_{2.5,\,corrected}$ = PM$_{2.5}$ × s$_1$ + T × s$_2$ + b | 0.33 | 1.8 | 0.48 | 1.7 | 0.52 | 1.6 | 0.53 | 1.6 |
| 4  | +D | PM$_{2.5,\,corrected}$ = PM$_{2.5}$ × s$_1$ + D × s$_2$ + b | 0.44 | 1.7 | 0.49 | 1.7 | 0.53 | 1.6 | 0.53 | 1.6 |
| 5  | +RH x T | PM$_{2.5,\,corrected}$ = PM$_{2.5}$ × s$_1$ + RH × s$_2$ + T × s$_3$ + RH × T × s$_4$ + b | 0.48 | 1.7 | 0.53 | 1.6 | 0.58 | 1.5 | 0.58 | 1.5 |
| 6  | +RH x D | PM$_{2.5,\,corrected}$ = PM$_{2.5}$ × s$_1$ + RH × s$_2$ + D × s$_3$ + RH × D × s$_4$ + b | 0.49 | 1.7 | 0.56 | 1.6 | 0.59 | 1.5 | 0.60 | 1.5 |
| 7  | +D x T | PM$_{2.5,\,corrected}$ = PM$_{2.5}$ × s$_1$ + D × s$_2$ + T × s$_3$ + D × T × s$_4$ + b | 0.47 | 1.7 | 0.53 | 1.6 | 0.56 | 1.5 | 0.57 | 1.5 |
| 8  | +RH x T x D | PM$_{2.5,\,corrected}$ = PM$_{2.5}$ × s$_1$ + RH × s$_2$ + T × s$_3$ + D × s$_4$ + RH × T × s$_5$ + RH × D × s$_6$ + T × D × s$_7$ + RH × T × D × s$_8$ + b | 0.55 | 1.6 | 0.61 | 1.5 | 0.66 | 1.4 | 0.66 | 1.4 |
| 9  | PM x RH | PM$_{2.5,\,corrected}$ = PM$_{2.5}$ × s$_1$ + | 0.43 | 1.7 | 0.52 | 1.6 | 0.54 | 1.6 | 0.58 | 1.5 |

| | | | | | | | | | |
|---|---|---|---|---|---|---|---|---|---|
| | | $RH \times s_2 + RH \times PM_{2.5} \times s_3 + b$ | | | | | | | |
| 10 | PM x D | $PM_{2.5, corrected} = PM_{2.5} \times s_1 + D \times s_2 + D \times PM_{2.5} \times s_3 + b$ | 0.50 | 1.7 | 0.56 | 1.58 | 0.63 | 1.4 | 0.63 | 1.4 |
| 11 | PM x T | $PM_{2.5, corrected} = PM_{2.5} \times s_1 + T \times s_2 + T \times PM_{2.5} \times s_3 + b$ | 0.34 | 1.8 | 0.51 | 1.6 | 0.52 | 1.6 | 0.58 | 1.5 |
| 12 | PM x nonlinear RH | $PM_{2.5, corrected} = PM_{2.5} \times s_1 + \frac{RH^2}{(1-RH)} \times s_2 + \frac{RH^2}{(1-RH)} \times PM_{2.5} \times s_3 + b$ | 0.43 | 1.7 | 0.59 | 1.5 | 0.63 | 1.4 | 0.68 | 1.3 |
| 13 | PM x RH x T | $PM_{2.5, corrected} = PM_{2.5} \times s_1 + RH \times s_2 + T \times s_3 + PM_{2.5} \times RH \times s_4 + PM_{2.5} \times T \times s_5 + RH \times T \times s_6 + PM_{2.5} \times RH \times T \times s_7 + b$ | 0.56 | 1.6 | 0.67 | 1.4 | 0.70 | 1.3 | 0.76 | 1.2 |
| 14 | PM x RH x D | $PM_{2.5, corrected} = PM_{2.5} \times s_1 + RH \times s_2 + D \times s_3 + PM_{2.5} \times RH \times s_4 + PM_{2.5} \times D \times s_5 + RH \times D \times s_6 + PM_{2.5} \times RH \times D \times s_7 + b$ | 0.54 | 1.6 | 0.66 | 1.4 | 0.68 | 1.4 | 0.76 | 1.2 |
| 15 | PM x T x D | $PM_{2.5, corrected} = PM_{2.5} \times s_1 + T \times s_2 + D \times s_3 + PM_{2.5} \times T \times s_4 + PM_{2.5} \times D \times s_5 + T \times D \times s_6 + PM_{2.5} \times T \times D \times s_7 + b$ | 0.55 | 1.6 | 0.66 | 1.4 | 0.68 | 1.4 | 0.76 | 1.2 |
| 16 | PM x RH x T x D | $PM_{2.5, corrected} = PM_{2.5} \times s_1 + RH \times s_2 + T \times s_3 + D \times s_4 + PM_{2.5} \times RH \times s_5 + PM_{2.5} \times T \times s_6 + T \times RH \times s_7 + PM_{2.5} \times D \times s_8 + D \times RH \times s_9 + D \times T \times s_{10} + PM_{2.5} \times RH \times T \times s_{11} + PM_{2.5} \times RH \times D \times s_{12} + PM_{2.5} \times D \times T \times s_{13} + D \times RH \times T \times s_{14} + PM_{2.5} \times RH \times T \times D \times s_{15} + b$ | 0.61 | 1.5 | 0.69 | 1.4 | 0.73 | 1.26 | 0.78 | 1.2 |
| | **Machine Learning (10-fold CV)** | | | | | | | | | |
| 17 | Random Forest | $PM_{2.5, corrected} = f(PM_{2.5}, T, RH)$ | 0.97 | 0.45 | 0.99 | 0.29 | 0.92 | 0.77 | 0.94 | 0.69 |

*Table S2*: Mobile monitoring summary data.

| Date | Sampling time (Number of measurements) | Sampling area | Vehicle Mean Speed (m/s) | PM$_{2.5}$ Mean (Median) µg/m³ ($\times$ 10) | | RH Mean (Median) (%) | | Temperature Mean (Median) ($^0$C) | | OPC position vis a vis vehicle direction | R (RMSE (µg/m³)) comparing corrected 5-second DustTrak and uncorrected OPC measurements | |
|---|---|---|---|---|---|---|---|---|---|---|---|---|
| | | | | OPC1 | OPC2 | OPC1 | OPC2 | OPC1 | OPC2 | | OPC 1 | OPC 2 |
| 02/11/202 | 09:18- | "South | 5.2 | 9 | - | 27.8 | - | 17.6 | - | OPC 1: flow rate parallel to | 0.31 | - |

| Date | Time (duration s) | Route | | | | | | | | Notes | | |
|---|---|---|---|---|---|---|---|---|---|---|---|---|
| 2 | 13:59 (3,351) | route": I-93 tunnels, Downtown Boston, South Boston, Dorchester, Roxbury, North Quincy | | (3) | | (27.6) | | (17.5) | | direction of vehicle 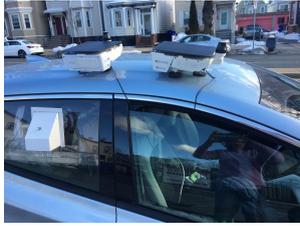 | (4.5) | |
| 2/16/2022 | 13:10-16:35 (2,458) | "North route": Everett, Revere, Chelsea, Winthrop, East Boston | 5.3 | 12 (5) | - | 40.0 (39.8) | - | 12.1 (12.0) | - | OPC 1: flow rate parallel to direction of vehicle | 0.37 (5.8) | - |
| 2/17/2022 | 09:15-13:42 (3,183) | South route | 5.5 | 21 (14) | - | 42.0 (42.0) | - | 22.8 (22.8) | - | OPC 1: flow rate parallel to direction of vehicle | 0.54 (5.4) | - |
| 2/23/2022 | 08:50-12:58 (2,976) | North route | 4.7 | 14 (12) | - | 55.5 (57.9) | - | 23.4 (22.5) | - | OPC 1: flow rate parallel to direction of vehicle 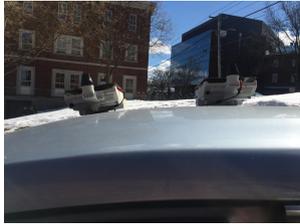 | 0.21 (6.3) | - |
| 2/24/2022 | 09:21-13:20 (2,869) | South route | 4.8 | 4 (2) | - | 26.8 (26.8) | - | 4.0 (3.6) | - | OPC 1: flow rate parallel to direction of vehicle | 0.29 (7.3) | - |
| 3/2/2022 | 13:33-18:26 (3,440) | South route + extra run through I-93 tunnels | 6.5 | 5 (2) | - | 34.2 (33.7) | - | 11.9 (11.9) | - | OPC 1: flow rate parallel to direction of vehicle | 0.25 (6.3) | - |
| 3/3/2022 | 13:39-16:23 (1,952) | North route | 6.6 | 5 (2) | - | 24.7 (24.1) | - | 10.6 (10.2) | - | OPC 1: flow rate parallel to direction of vehicle 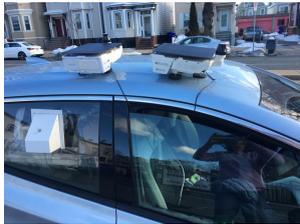 | 0.22 (3.4) | - |
| 3/8/2022 | 16:16-20:05 (2,707) | North route | 3.3 | 2 (1) | - | 24.5 (24.3) | - | 11.8 (11.9) | - | OPC 1: flow rate parallel to direction of vehicle | 0.27 (5.6) | - |
| 3/10/2022 | 12:22-16:52 (3,220) | South route | 5.5 | 9 (3) | - | 31.1 (30.7) | - | 16.5 (16.6) | - | OPC 1: flow rate perpendicular to direction of vehicle | 0.24 (8.3) | - |

| Date | Time | Route | | | | | | | | | | |
|---|---|---|---|---|---|---|---|---|---|---|---|---|
| | | | | | | | | | | 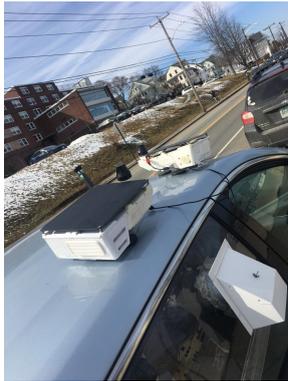 | | |
| 3/23/2022 | 11:29 - 14:40 (2,298) | South route | 5.6 | 2 (1) | 5 (3) | 29.5 (29.2) | 23.3 (23.8) | 15.5 (15.7) | 16.7 (16.6) | OPC 1: flow rate perpendicular to direction of vehicle OPC 2: flow rate parallel to the direction of the vehicle | 0.15 (6.0) | 0.45 (5.6) |
| 3/30/2022 | 12:06 - 16:02 (2,828) | South route | 6.1 | 2 (1) | 9 (5) | 19.2 (18.9) | 13.3 (12.7) | 16.6 (16.8) | 18.2 (18.6) | OPC 1: flow rate perpendicular to direction of vehicle OPC 2: flow rate parallel to the direction of the vehicle | 0.33 (4.7) | 0.44 (4.1) |
| 4/05/2022 | 10:27 - 14:24 (2,845) | South route | 5.4 | 4 (3) | 7 (4) | 22.6 (22.9) | 16.4 (16.4) | 22.8 (22.1) | 24.3 (23.8) | OPC 1: flow rate perpendicular to direction of vehicle OPC 2: flow rate parallel to the direction of the vehicle | 0.33 (8.9) | 0.24 (8.8) |

*Note that due to the low uncorrected OPC PM$_{2.5}$ measurements, we report PM$_{2.5}$ concentrations × 10.*

*Table S3*: Comparing corrected OPC measurements using different calibration models derived from minute- and hourly-averaged collocated data with that of the corresponding corrected (from the calibration model developed using hourly-averaged collocation data) DustTrak data in our mobile deployment. The R and RMSE when comparing the corrected DustTrak measurements using minute-averaged and hourly-averaged data during the collocation were 1 and 0.69 μg/m³, respectively.

| ID | Name | Model | Calibration derived from minute-averaged data | | | | Calibration derived from hourly-averaged data | | | |
|---|---|---|---|---|---|---|---|---|---|---|
| | | | DustTrak and OPC 1 | | DustTrak and OPC 2 | | DustTrak and OPC 1 | | DustTrak and OPC 2 | |
| | | | R | RMSE (μg/m³) | R | RMSE (μg/m³) | R | RMSE (μg/m³) | R | RMSE (μg/m³) |
| | **Raw OPC measurements** | | | | | | | | | |
| 0 | Raw | | 0.26 | 6.5 | 0.20 | 6.9 | 0.26 | 6.5 | 0.20 | 6.9 |
| | **Multivariate Regression** | | | | | | | | | |
| 1 | Linear | PM$_{2.5, corrected}$ = PM$_{2.5}$ × s1 + b | 0.26 | 7.0 | 0.20 | 4.0 | 0.26 | 16.8 | 0.20 | 4.0 |
| 2 | +RH | PM$_{2.5, corrected}$ = PM$_{2.5}$ × s$_1$ + RH × s$_2$ + b | 0.26 | 5.9 | 0.20 | 4.0 | 0.26 | 16.6 | 0.20 | 4.0 |
| 3 | +T | PM$_{2.5, corrected}$ = PM$_{2.5}$ × s$_1$ + T × s$_2$ + b | 0.26 | 7.3 | 0.31 | 3.9 | 0.26 | 17.6 | 0.31 | 3.9 |

| | | | | | | | | | |
|---|---|---|---|---|---|---|---|---|---|
| 4 | +D | PM$_{2.5, corrected}$ = PM$_{2.5}$ × s$_1$ + D × s$_2$ + b | 0.29 | 5.2 | 0.42 | 4.2 | 0.27 | 12.6 | 0.39 | 4.3 |
| 5 | +RH x T | PM$_{2.5, corrected}$ = PM$_{2.5}$ × s$_1$ + RH × s$_2$ + T × s$_3$ + RH × T × s$_4$ + b | 0.27 | 5.5 | 0.33 | 4.1 | 0.26 | 14.9 | 0.31 | 4.1 |
| 6 | +RH x D | PM$_{2.5, corrected}$ = PM$_{2.5}$ × s$_1$ + RH × s$_2$ + D × s$_3$ + RH × D × s$_4$ + b | 0.29 | 5.7 | 0.59 | 4.4 | 0.27 | 15.6 | 0.56 | 4.4 |
| 7 | +D x T | PM$_{2.5, corrected}$ = PM$_{2.5}$ × s$_1$ + D × s$_2$ + T × s$_3$ + D × T × s$_4$ + b | 0.28 | 5.4 | 0.48 | 4.2 | 0.27 | 13.8 | 0.45 | 4.2 |
| 8 | +RH x T x D | PM$_{2.5, corrected}$ = PM$_{2.5}$ × s$_1$ + RH × s$_2$ + T × s$_3$ + D × s$_4$ + RH × T × s$_5$ + RH × D x s$_6$ + T × D × s$_7$ + RH × T × D × s$_8$ + b | 0.25 | 5.9 | 0.11 | 4.3 | 0.26 | 16.3 | 0.08 | 4.4 |
| 9 | PM x RH | PM$_{2.5, corrected}$ = PM$_{2.5}$ × s$_1$ + RH × s$_2$ + RH × PM$_{2.5}$ × s$_3$ + b | 0.28 | 5.6 | 0.31 | 4.1 | 0.26 | 14.8 | 0.26 | 4.3 |
| 10 | PM x D | PM$_{2.5, corrected}$ = PM$_{2.5}$ × s$_1$ + D × s$_2$ + D × PM$_{2.5}$ × s$_3$ + b | 0.17 | 18.4 | 0.28 | 4.9 | 0.19 | 36.4 | 0.27 | 5.4 |
| 11 | PM x T | PM$_{2.5, corrected}$ = PM$_{2.5}$ × s$_1$ + T × s$_2$ + T × PM$_{2.5}$ × s$_3$ + b | 0.25 | 8.2 | 0.26 | 4.1 | 0.26 | 17.9 | 0.24 | 4.3 |
| 12 | PM x nonlinear RH | PM$_{2.5, corrected}$ = PM$_{2.5}$ × s$_1$ + $\frac{RH^2}{(1-RH)}$ × s$_2$ + $\frac{RH^2}{(1-RH)}$ × PM$_{2.5}$ × s$_3$ + b | 0.22 | 19.2 | 0.20 | 4.8 | 0.23 | 40.3 | 0.20 | 5.3 |
| 13 | PM x RH x T | PM$_{2.5, corrected}$ = PM$_{2.5}$ × s$_1$ + RH × s$_2$ + T × s$_3$ + PM$_{2.5}$ × RH × s$_4$ + PM$_{2.5}$ × T × s$_5$ + RH × T × s$_6$ + PM$_{2.5}$ × RH × T × s$_7$ + b | 0.18 | 23.9 | 0.27 | 5.6 | 0.20 | 44.1 | 0.25 | 7.1 |
| 14 | PM x RH x D | PM$_{2.5, corrected}$ = PM$_{2.5}$ × s$_1$ + RH × s$_2$ + D × s$_3$ + PM$_{2.5}$ × RH × s$_4$ + PM$_{2.5}$ × D × s$_5$ + RH × D × s$_6$ + PM$_{2.5}$ × RH × D × s$_7$ + b | 0.21 | 17.3 | 0.28 | 4.8 | 0.21 | 42.2 | 0.23 | 7.1 |
| 15 | PM x T x D | PM$_{2.5, corrected}$ = PM$_{2.5}$ × s$_1$ + T × s$_2$ + D × s$_3$ + PM$_{2.5}$ × T × s$_4$ + PM$_{2.5}$ × D × s$_5$ + T × D × s$_6$ + PM$_{2.5}$ × T × D × s$_7$ + b | 0.17 | 43.5 | 0.29 | 5.9 | 0.19 | 55.9 | 0.27 | 8.0 |
| 16 | PM x RH x T x D | $PM_{2.5, corrected} = PM_{2.5} \times s_1 + RH \times s_2 + T \times s_3 + D \times s_4 + PM_{2.5} \times RH \times s_5 + PM_{2.5} \times T \times s_6 + T \times RH \times s_7 + PM_{2.5} \times D \times s_8 + D \times RH \times s_9 + D \times T \times s_{10} + PM_{2.5} \times RH \times T \times s_{11} + PM_{2.5} \times RH \times D \times s_{12} + PM_{2.5} \times D \times T \times s_{13} + D \times RH \times T \times s_{14} + PM_{2.5} \times RH \times T \times D \times s_{15} + b$ | 0.20 | 33.1 | 0.17 | 5.7 | 0.11 | 101.2 | 0.19 | 8.9 |
| | **Machine Learning (10-fold CV)** | | | | | | | | | |
| 17 | Random Forest | $PM_{2.5, corrected} = f(PM_{2.5}, T, RH)$ | 0.17 | 4.1 | 0.16 | 4.6 | 0.25 | 3.3 | 0.17 | 4.5 |

*Table S4*: Sum of Squares (Explained variation (%) in the difference and absolute difference between uncorrected and corrected OPC measurements, and corrected OPC and DustTrak $PM_{2.5}$ measurements) from an ANOVA analysis. The T and RH used in ANOVA analyses comparing the OPC measurements came from City Scanner corresponding to OPC 1. When comparing OPC 1 and OPC 2 measurements with DustTrak data, we used T and RH data corresponding to the City Scanner corresponding to OPC 1 and 2, respectively.

| | OPC | No | Orientation | RH | T | Speed | Sky view factor | Day | Hour | Road Class | Residuals | Total Sum of Squares |
|---|---|---|---|---|---|---|---|---|---|---|---|---|
| **Absolute difference between the uncorrected OPC $PM_{2.5}$ measurements** | | 1 | - | 0 (0.00%) | 46* (0.25%) | 0 (0.00%) | 0 (0.00%) | 151* (0.81%) | 139* (0.75%) | 35* (0.19%) | 18,195 (98.00%) | 18,566 |
| **Absolute difference between the corrected OPC $PM_{2.5}$ measurements using Model 6 (derived from minute-averaged collocated measurements)** | | 3 | - | 6* (0.07%) | 1,111* (12.20%) | 0 (0.00%) | 6* (0.07%) | 57* (0.63%) | 398* (4.37%) | 278* (3.05%) | 7,252 979.62%) | 9,108 |
| **Absolute difference between the corrected DustTrak measurements and the OPC $PM_{2.5}$ measurements** | 1 | 5 | 115* (0.02%) | 1,091* (0.18%) | 74 (0.01%) | 461* (0.07%) | 2,254* (0.36%) | 40,990* (6.63%) | 8,161* (1.32%) | 1,755* (0.28%) | 563,533 (91.12%) | 618,434 |
| | 2 | 6 | - | 602* (1.72%) | 13,232* (37.74%) | 39* (0.11%) | 37* (0.11%) | 1,009* (2.88%) | 2,468* (7.04%) | 195* (0.56%) | 17,478 (49.85%) | 35,060 |
| **Difference between the uncorrected OPC $PM_{2.5}$ measurements** | | 9 | - | 7 (0.036%) | 330* (1.69%) | 0 (0.00%) | 0 (0.00%) | 198* (1.01%) | 59* (0.30%) | 20 (0.10%) | 18,953 (96.86%) | 19,567 |
| **Difference between the corrected OPC $PM_{2.5}$** | | 11 | - | 9* (0.08%) | 1,832* (16.64%) | 1 (0.01%) | 8* (0.07%) | 229* (2.08%) | 210* (1.91%) | 152* (1.38%) | 8,568 (77.83%) | 11,009 |
| measurements using Model 6 (derived from minute-averaged collocated measurements) | | | | | | | | | | | 43 | |
| **Difference between the corrected DustTrak measurements and the OPC PM$_{2.5}$ measurements** | 1 | 13 | 1,091* (0.12%) | 123,356* (13.45%) | 39,338* (4.29%) | 1 (0.00%) | 128* (0.01%) | 96,875* (10.56%) | 6,209* (0.68%) | 1,482* (0.16%) | 648,465 (70.72%) | 916,945 |
| | 2 | 14 | - | 587* (1.61%) | 13,231* (36.34%) | 43* (0.12%) | 44* (0.12%) | 892* (2.45%) | 2,280* (6.26%) | 182* (0.50%) | 19,147 (52.59%) | 36,406 |